\documentclass[prb,eqsecnum,twocolumn,floatfix,showpacs]{revtex4}
\usepackage{graphicx}% Include figure files

\def\fmfL(#1,#2,#3)#4{\put(#1,#2){\makebox(0,0)[#3]{#4}}}

\newcommand{\Ham}{\mathcal{H}}
\newcommand{\Act}{\mathcal{S}}

\newcommand{\Mean}[1]{\left\langle#1\right\rangle}
\newcommand{\sub}[1]{_{\mathrm{#1}}}
\newcommand{\Dim}[1]{\dim [#1]}
\newcommand{\units}[1]{\ensuremath{\,\mathrm{#1}}}

\newcommand{\be}{\begin{equation}}
\newcommand{\ee}{\end{equation}}
\newcommand{\bea}{\begin{eqnarray}}
\newcommand{\eea}{\end{eqnarray}}
\newcommand{\punc}[1]{\;\mathrm{#1}}

\newcommand{\ie}{{\it i.e.}}

\newcommand{\mathsmall}{\textstyle}
\newcommand{\smallfrac}[2]{{\mathsmall\frac{#1}{#2}}}
\newcommand{\half}{\smallfrac{1}{2}}

\newcommand{\Order}[1]{\mathcal{O}\left(#1\right)}
\newcommand{\parder}[2]{\frac{\partial #1}{\partial #2}}
\newcommand{\parderat}[3]{{\left(\frac{\partial #1}{\partial #2}\right)}_{#3}}

\newcommand{\Gt}{\widetilde{G}} % Effective propagators for b and \psi
\newcommand{\Xit}{\widetilde{\Xi}}
\newcommand{\xit}{\tilde{\xi}}
\newcommand{\DeltaLDA}{\widetilde{\Delta}}
\newcommand{\lamt}{\tilde{\lambda}}
\newcommand{\LW}{\Phi\sub{LW}}
\begin{document}
\setlength{\unitlength}{1mm}

\title{Depletion of the Bose-Einstein condensate in Bose-Fermi mixtures}

\author{Stephen Powell}
\author{Subir Sachdev}
\affiliation{Department of Physics, Yale University, P.O. Box
208120, New Haven, CT 06520-8120}
\author{Hans Peter B\"uchler}
\affiliation{Institut f\"ur Theoretische Physik, Universit\"at Innsbruck,
Technikerstra\ss e 25, A-6020 Innsbruck, Austria}

\begin{abstract}
We describe the properties of a mixture of fermionic and bosonic
atoms, as they are tuned across a Feshbach resonance associated
with a fermionic molecular state. Provided the number of fermionic
atoms exceeds the number of bosonic atoms, we argue that there is
a critical detuning at which the Bose-Einstein condensate (BEC) is
completely depleted. The phases on either side of this quantum
phase transition can also be distinguished by the distinct
Luttinger constraints on their Fermi surfaces. In both phases, the
total volume enclosed by all Fermi surfaces is constrained by the
total number of fermions. However, in the phase without the BEC,
which has two Fermi surfaces, there is a {\em second\/} Luttinger
constraint: the volume enclosed by one of the Fermi surfaces is
constrained by the total number of {\em bosons}, so that the
volumes enclosed by the two Fermi surfaces are separately
conserved. The phase with the BEC may have one or two Fermi
surfaces, but only their total volume is conserved. We obtain the
phase diagram as a function of atomic parameters and temperature,
and describe critical fluctuations in the vicinity of all
transitions. We make quantitative predictions valid for the case
of a narrow Feshbach resonance, but we expect the qualitative
features we describe to be more generally applicable.
As an aside, we point out intriguing connections between the BEC
depletion transition and the transition to the fractionalized Fermi
liquid in Kondo lattice models.
\end{abstract}

\vspace{3cm}

\maketitle
\section{Introduction}
\label{sec:intro}

The Feshbach resonance has emerged as a powerful tool in studying
ultracold atoms in regimes of strong interactions. For two
isolated atoms scattering off each other, the Feshbach resonance
is a singularity in their scattering length due to the coupling
of the atomic states to a molecular bound state.
\cite{stwalley1976,tiesinga1993} The singularity (at $\nu = 0$)
occurs as a function of the detuning $\nu$, which is a measure of
the energy difference between the atomic and molecular states. The
value of $\nu$ can be varied by an applied magnetic field, and
this effectively allows one to tune the strength of the atomic
interactions.

For systems in the thermodynamic limit, with a finite density of
atoms, there is no singularity at $\nu=0$. Nevertheless, the
vicinity of $\nu=0$ is a regime of interesting many body effects.
For a Feshbach resonance between two identical fermionic atoms,
the many body ground state changes from a Bose-Einstein condensate
(BEC) of molecules ($\nu \ll 0$) to a Bardeen-Cooper-Schrieffer
(BCS) superfluid descended from a Fermi gas of atoms ($\nu \gg
0$). It is important to note that there is no true fundamental
distinction between the BEC and BCS states here, and so the two
limits are connected by a smooth crossover. Recent experiments
\cite{jochim2003,greiner2003,zwierlein2003}  on $^{6}$Li and
$^{40}$K atoms have succeeded in observing the BEC of molecules.

The consequences of the two-body Feshbach resonance are very
different for other atomic statistics. For a Feshbach resonance
between two identical bosonic atoms, it has been argued recently
\cite{leo,ss} that there is indeed a sharp singularity, \ie\ a
quantum phase transition, in the many body system as a
function of $\nu$. This singularity is not precisely at $\nu=0$,
but is shifted away from it; it is not directly a reflection of
the singularity in the scattering length of two isolated atoms,
but is a new many body effect. Here, the two limiting states are a
BEC of molecules ($\nu \ll 0$) and a BEC of atoms ($\nu \gg 0$).
Unlike the fermionic case above, these two states cannot be
connected smoothly to each other. The fundamental distinction
between these states becomes apparent upon examining the quantum
numbers of the vortices in the condensate: the quantum of
circulation differs by a factor of 2 in the two limits, being
determined respectively by the mass of a molecule or of an atom.

In the present paper, we will consider the remaining case of a
mixture of two distinct types of atoms, one fermionic and the
other bosonic. Mixtures of fermionic $^{6}$Li and bosonic $^{7}$Li
atoms were studied by Truscott {\em et al.} \cite{truscott} and
Schreck {\em et al.},\cite{schreck} and they succeeded in
achieving simultaneous quantum degeneracy in both species of
atoms. Recently, Feshbach resonances have been observed between
bosonic $^{23}$Na and fermionic $^{6}$Li atoms by Stan {\em et
al.}, \cite{stan} and between bosonic $^{87}$Rb and fermionic
$^{40}$K atoms by Inouye {\em et al.} \cite{inouye} So the time
is clearly appropriate to examine the many body properties of such
mixed Bose and Fermi gases across a Feshbach resonance, in which
the fermionic molecule of the two unlike atoms can also reach
quantum degeneracy.

Our primary result is that such a mixture of fermionic and bosonic
atoms also has a quantum phase transition. Again this transition
is a many body effect, and does not occur precisely at $\nu=0$. We
will map out the phase diagram as functions of $\nu$, temperature
($T$), and the densities of the atoms (see Figs.~\ref{Graph0},
\ref{Phases}, \ref{ZeroTphases}, \ref{ZeroTmu}) and also describe
the strong consequences of thermal and quantum fluctuations in
the vicinity of the phase transition. It is also possible that
the mixture phase separates: this will be studied in the Appendix,
where we determine the region of instability to phase separation
(see Fig.~\ref{ZeroTstability}).

As in the Fermi-Fermi and Bose-Bose cases, the existence of a
quantum phase transition for the Bose-Fermi case can be easily
understood by characterizing the two limiting cases. For $\nu \gg
0$, there is a BEC of the bosonic atoms and a Fermi surface of the
fermionic atoms. In contrast, for $\nu \ll 0$, there is a Fermi
surface of the fermionic molecules. If the number of the fermionic
($N_f$) and bosonic ($N_b$) atoms are unequal to each other, for $
\nu \ll 0$, there will also be some residual atoms which are not
in molecules forming their own ground state: for $N_f > N_b$ the
extra fermions will form a separate Fermi surface of atoms, while
for $N_b > N_f$, the extra bosons will form an atomic BEC. We note
that for $N_f > N_b$, scanning the detuning $\nu$ takes us between
limits with and without an atomic BEC. Consequently there must be
a critical detuning at which the atomic BEC is completely
depleted, and all the bosonic atoms have been absorbed into
molecules.

A novel feature of this quantum phase transition is that it can be
entirely characterized in terms of the Luttinger constraints on
the Fermi surfaces.
\renewcommand{\labelenumi}{({\it\theenumi})}
\renewcommand{\theenumi}{\roman{enumi}}
\begin{enumerate}
\item Consider, first, the phase without the BEC with $\nu
\ll 0$. Here, there are 2 Fermi surfaces, one with Fermi
surface excitations which are primarily the fermionic atoms, while
the other has Fermi surface excitations which are primarily the
fermionic molecules. We establish in Section~\ref{luttinger} that
this phase obeys {\em two} Luttinger theorems: the atomic Fermi
surface encloses a volume associated with precisely $N_f-N_b$
states, while the molecular Fermi surface encloses precisely $N_b$
states.
\item Now consider the phase with the BEC. When the BEC is
small, this phase retains 2 Fermi surfaces, one primarily atomic
and the other primarily molecular. However, now the volumes
enclosed by these Fermi surfaces are not separately conserved;
only the total volume enclosed by both Fermi surfaces is required
to contain $N_f$ states. Eventually, for $\nu \gg 0$, the
molecular Fermi surface disappears entirely, and only a single
Fermi surface with $N_f$ states remains. The disappearance of the
molecular Fermi surface (in the presence of a BEC) is a second
quantum transition whose character we will also discuss briefly in
Section~\ref{EffActBoson}.
\end{enumerate}

This paper will determine the value of the critical $\nu$ for the
BEC depletion transition, and describe critical fluctuations in
its vicinity. At $T=0$, we will find in Section~\ref{EffActBoson}
that this critical point is generically in the universality class
\cite{fwgf,book} of the density-driven superfluid-insulator
transition with dynamic exponent $z=2$. There is also an
interesting quantum multi-critical point for $N_f = 2 N_b$ at
which the BEC depletion quantum transition has a different
character: this we will also describe. At $T>0$, the BEC depletion
transition is in the universality class of the
$\lambda$-transition of $^4$He, and so will display similar
critical singularities: a peak in the specific heat, and anomalies
in transport coefficients.

Our quantitative results are determined within a mean-field
picture, whose applicability is restricted to the case of a
`narrow' Feshbach resonance, where the relevant coupling is
sufficiently weak. We nonetheless expect our results to be
at least qualitatively applicable to the (experimentally more
common) `wide' Feshbach resonance.

We also find an additional $T=0$ quantum phase transition
involving the disappearance of the molecular Fermi surface. As
shown in Section~\ref{EffActBoson}, this is described by a $z=2$
critical theory of free fermions.

While our work was in progress, we learnt of the work of Yabu {\em
et al.} \cite{Yabu} who addressed some related issues, but only in
the limit of infinitesimal coupling between the atomic and
molecular degrees of freedom. We will note their limiting results
in Section~\ref{Zeroglimit}.

We now outline the contents of the body of the paper.

First, in Section \ref{BasicDefinitions}, we define the model
Hamiltonian that will be used throughout the rest of the paper. In
Section \ref{Zeroglimit}, we consider the limit of vanishing
coupling, where a purely classical analysis can be used.
\cite{Yabu}

Section \ref{MFforMu} finds the phase structure for finite coupling,
treating quantum-mechanical effects using a mean-field approach. In Section
\ref{luttinger}, we describe our results regarding Luttinger's theorem for
the system. In Section \ref{EffActBoson}, the mean-field result of
Section \ref{MFforMu} is reproduced using a field-theoretical
approach, which further allows us to characterize the critical
properties of the transition.

In Sections \ref{GaussianCorrections} and \ref{LowDensityApproximation}, two
corrections are calculated to the mean-field theory, which can be used to
determine the validity of this approximation. In Section
\ref{GaussianCorrections}, the two-loop corrections to the free energy are
found, while in Section \ref{LowDensityApproximation}, higher orders in the
coupling are included, within a low-density approximation.

In the Appendix, we consider the stability of the system against separation
into two regions with differing densities. It is shown that the system is indeed
stable for a broad range of parameters.

\section{Basic definitions}
\label{BasicDefinitions}

The system consists of bosonic atoms $b$ and fermionic atoms $f$ which combine
to form fermionic molecules $\psi$. The energy, relative to the chemical
potential $\mu$, is for the atoms
\be
\label{Definexif}
\xi^f_k=\epsilon^f_k-\mu^f=\frac{k^2}{2m^f} - \mu^f
\ee
\be
\xi^b_k=\epsilon^b_k-\mu^b=\frac{k^2}{2m^b} - \mu^b
\ee
and for the molecule
\be
\label{Definexipsi}
\xi^\psi_k=\epsilon^\psi_k-\mu^\psi=\frac{k^2}{2m^\psi} - \mu^\psi + \nu\punc{,}
\ee
including the detuning $\nu$. The masses obey $m^\psi=m^f+m^b$ and, because of
the interaction, the chemical potentials are related by $\mu^\psi=\mu^f+\mu^b$.

The grand Hamiltonian is
\bea
\Ham &=& \int_k (\xi^f_k f_k^\dag f_k
+ \xi^b_k b_k^\dag b_k + \xi^\psi_k \psi_k^\dag \psi_k)
\nonumber\\
&&-\: g\int_{k,k'}(\psi_{k+k'}^\dag f_k b_{k'} + b_{k'}^\dag f_k^\dag \psi_{k+k'})
\nonumber\\
&&+ \lambda\int_{k,k',\ell} b_{k+\ell}^\dag b_{k'-\ell}^\dag b_{k'}\, b_k\punc{,}
\label{Hamiltonian}
\eea
where $\int_k$ denotes $\int d^3 k\,/(2\pi)^3$.

We assume throughout that the fermion spin is
polarized along some direction, so that both $f$ and $\psi$ are treated as
spinless. The fourth term (in $g$) causes the bosonic and fermionic atoms to
couple and form molecules, while the final term (in $\lambda$) is an
interaction between pairs of bosons. We omit the interaction between fermions
because the exclusion principle forbids $s$-wave scattering between identical
fermions and we assume that the interaction between $f$ and $\psi$ will be
less important than the coupling $g$.

Taking the dimensions of momentum and energy to be unity, $\Dim{k} = \Dim{E} =
1$, we have $\Dim{\psi} = -\frac{3}{2}$ and the same for the operators $b$ and
$f$. (Throughout, we shall measure temperature, energy and frequency in the same
units, so that $\hbar = k\sub{B} = 1$.) The coupling constants
have dimensions $\Dim{g}=-\frac{1}{2}$ and $\Dim{\lambda}=-2$.

At temperature $T = 1/\beta\neq 0$, we have six dimensionless parameters.
First let $N_b$ be the total density of bosonic atoms, including those
bound in molecules, and let $N_f$ be the same for fermionic atoms. (We
consider a unit volume, so that density is synonymous with number.) In the
absence of any fermions, the bosons would condense at a temperature
\be
T_0 = \frac{2\pi}{m^b}{\left[\frac{N_b}{\zeta(\smallfrac{3}{2})}\right]}
^\frac{2}{3}\punc{.}
\ee
We can take as dimensionless parameters $T/T_0 = \beta_0/\beta$, $N_f/N_b$,
$m^f/m^b$, $\nu/T_0$, $\gamma^2/T_0$ and $\lambda^2 (m^b)^3 T_0$, where
\be
\label{Definegamma}
\gamma = \frac{g^2}{8\pi} {\left(\frac{2m^f m^b}{m^\psi}\right)}^{3/2}\punc{.}
\ee

In what follows, it will not usually be necessary to take account of the
coupling between bosons given by the final term of (\ref{Hamiltonian}).
Except within the condensed phase, which will be treated in Section
\ref{MFforMu}, the only effect of $\lambda$ is a renormalization of the boson
mass, which we assume has already been incorporated into the definition of
$m^b$.

\subsection{Physical units}

In order to relate these parameters to experimental values, we may choose a unit
of volume of $10^{-15}\units{cm^3}$, which gives the unit of momentum as roughly
$10^{-27}\units{kg\cdot m/s}$. Taking the unit of mass to be $6\units{amu}$,
corresponding to a lithium-6 atom, the unit of energy is roughly $7\times 10^{-
10}\units{eV}$ or $8\units{\mu K}$.

For a Feshbach resonance, we assume the expression \cite{Duine}
\be
g = \sqrt{\frac{2\pi a\sub{bg}\,\Delta B \Delta\mu}{m}}\punc{,}
\ee
where $a\sub{bg}$ is the background scattering length, $\Delta B$ is the width
of the resonance and $\Delta\mu$ is the difference in magnetic moments.
Using the observed background scattering length between lithium-6 and -7 of
$a\sub{bg} = 2.0\units{nm}$,\cite{schreck} we may estimate the coupling constant.
Taking, for instance $\Delta B = 1\units{G}$, $\Delta \mu = \mu\sub{B}$, the
Bohr magneton, we find $g \simeq 1$ in our units. For a boson density
$N_b = 10^{15}\units{cm^{-3}}$ and mass $m_b = m_f = 6\units{amu}$, the value
$g = 1$ gives a dimensionless coupling of $\gamma^2/T_0 = 5\times 10^{-4}$.

While the width of the resonance used here, $\Delta B = 1\units{G}$, is
sufficiently large that $\Delta\mu \Delta B \gtrsim T_0$, it is nonetheless
somewhat smaller than typical experimental values. For our purposes, a more
relevant measure of the resonance `width' is the
lifetime of the molecule state in the vacuum (for $\nu > 0$). This is calculated
in Section \ref{LowDensityApproximation}, where we show that it is determined by
the constant $\gamma$. Since the relevant energies are on the order of $T_0$,
the condition for a narrow resonance is that $\gamma^2/T_0 \ll 1$. For the
numerical results throughout this paper, we will always remain in this narrow
limit, which is analytically more accessible. As noted above, we expect our
results to be at least qualitatively applicable even for the wider Feshbach
resonances observed experimentally.

Following Ref.~\onlinecite{ss}, we take
\be
\lambda = \frac{2\pi}{m^b}a_{bb}\punc{,}
\ee
where for $a_{bb}$, the scattering length for the boson-boson interaction,
we use $a_{bb} = 0.27 \units{nm}$,\cite{schreck} giving
\be
\lambda^2 (m^b)^3 T_0 = 2\times 10^{-3}\punc{.}
\ee

The detuning $\nu$ appearing in the molecular
dispersion relation (\ref{Definexipsi}) is given by \cite{Duine}
\be
\nu = \Delta\mu(B - B_0)\punc{,}
\ee
where $B_0$ is the magnetic field at resonance and $B$ is the applied field.

\section{The limit $g\rightarrow 0$}
\label{Zeroglimit}

The case of vanishing coupling, which can be addressed with a classical
approach, has been considered by Yabu {\em et al.} \cite{Yabu} (The results
presented in this section produce Fig.~3 of Ref.~\onlinecite{Yabu}, which corresponds
to our Fig.~\ref{ZeroTphases}, below.)

For simplicity, we restrict the analysis to zero temperature, but similar
arguments can be made for nonzero temperatures. We call the two Fermi energies
$\epsilon^f_0$ and $\epsilon^\psi_0$, and the corresponding wavenumbers $k^f_0$
and $k^\psi_0$. At zero temperature, all bosons are at $\epsilon^b = 0$ and
fermionic atoms or molecules must be added at their respective Fermi levels.

The atomic Fermi surface (FS) vanishes when all the fermionic atoms are contained
in molecules, so that
\be
k_0^\psi = (6\pi^2\,N_f)^{1/3}\punc{.}
\ee
(The number of states within a unit sphere in momentum space is $1/6\pi^2$.)
For this arrangement to be favorable energetically, the molecular Fermi energy,
$\epsilon_0^\psi$, must remain below the lowest atomic energy level. The boundary
of the phase without an atomic FS is therefore where
\be
\frac{1}{2m^\psi}(N_f)^{2/3} + \frac{\nu}{(6\pi^2)^{2/3}} = 0\punc{.}
\ee
Similarly, the molecular FS vanishes at the point when
\be
\frac{1}{2m^f}(N_f)^{2/3} - \frac{\nu}{(6\pi^2)^{2/3}} = 0\punc{.}
\ee
The atomic (molecular) FS is therefore only absent for negative (positive)
detuning $\nu$.

To find the boundary of the phase with a BEC, we must consider the depletion of
the condensate. Bosons will take fermions and form molecules as long as their
final energy is lower, \ie\ $\epsilon^\psi_0 < \epsilon^f_0 + 0$.
The phase boundary is therefore where $\epsilon^f_0 = \epsilon^\psi_0$, which gives
\be
\frac{1}{2m^f}{\left({N_f - N_b}\right)}^{2/3} -
\frac{1}{2m^\psi}{\left({N_b}\right)}^{2/3}
= \frac{\nu}{(6\pi^2)^{2/3}}\punc{,}
\ee
where the wavenumbers have been determined from $N_b$ and $N_f$, using the fact
that there is no condensate.

It should be noted that, in this limit, the
coupling to fermionic atoms reduces the tendency of the bosons to condense.
(The same is true at nonzero temperature.)

\section{Mean-field theory}
\label{MFforMu}

It is possible to go beyond the classical analysis used for vanishing coupling,
by using mean-field theory. We will present here two parallel developments, in
this section and Section \ref{EffActBoson}, respectively. The first is based on
single-particle quantum mechanics, using the mixing between the fermionic
dispersion relations caused by the presence of a BEC. The
second uses a field-theoretic approach and considers perturbative corrections to
the bosonic propagator. The former has the advantage of giving a somewhat
clearer physical picture and leading more directly to thermodynamic results
(such as the question of phase separation, considered in the Appendix), while
the latter leads naturally to higher-order corrections.

In the remainder of this section, we present the quantum-mechanical approach,
starting from the Hamiltonian (\ref{Hamiltonian}). First, in Section
\ref{MFHamiltonian}, we make a mean-field approximation and diagonalize the new
Hamiltonian. We then find the condition that a BEC should
be energetically favorable, within this approximation.

Since the Hamiltonian is defined in the grand canonical ensemble, we must then
relate the chemical potentials to the particle numbers, in Section
\ref{MFFixedN}. Within the mean-field approximation, it is sufficient to find
this relation to order zero in the coupling, neglecting two-loop corrections to
the free energy.\cite{ohashi:130402} Later, in Section
\ref{GaussianCorrections}, we determine the higher-order corrections.

In Section \ref{ZeroT}, we restrict our attention to the case of zero temperature,
where transitions occur between states with different numbers of Fermi surfaces. We
identify the positions of these transitions and present the full phase diagram
for $T = 0$.

\subsection{Mean-field Hamiltonian}
\label{MFHamiltonian}

Replacing the boson field $b_k$ in (\ref{Hamiltonian}) by a real constant
$\varphi$ gives
\bea
\Ham\sub{mf} &=& \int\!\!\frac{d^3 k}{(2\pi)^3}\!\left[ \xi^f_k f_k^\dag f_k
+ \xi^\psi_k \psi_k^\dag \psi_k
- g\varphi(\psi_k^\dag f_k + f_k^\dag \psi_k) \right]\nonumber\\
&&- \:\mu^b \varphi^2 + \lambda \varphi^4\punc{,}
\label{HamBefore}
\eea
which can be diagonalized to
\be
\label{HamAfter}
\Ham\sub{mf} = \int\frac{d^3 k}{(2\pi)^3} \left( \xi^F_k F_k^\dag F_k
+ \xi^\Psi_k \Psi_k^\dag \Psi_k \right) - \mu^b \varphi^2
+ \lambda \varphi^4\punc{.}
\ee
The dispersion relations for the mixed fermions $F$, $\Psi$ are
\be
\xi^{F,\Psi}_k = \half\left( \xi^f_k + \xi^\psi_k \right) \pm
\half\sqrt{{\left(\xi^f_k-\xi^\psi_k\right)}^2 + 4g^2\varphi^2}
\label{MixedDispersion}\punc{,}
\ee
with the choice that $\xi^F_k \ge \xi^\Psi_k$ for all $k$.

Since the mixing will cause the dispersion relations to separate, the total
energy of the fermions is lowered by nonzero $\varphi$. This quantum-mechanical
effect, in contrast to the purely classical effect described in Section
\ref{Zeroglimit}, therefore acts to favor condensation.

We must analyze the energetics to determine the point at which a condensate
becomes favorable. The grand free energy $\Phi$ is minimized at temperature
$1/\beta$ by a Fermi-Dirac distribution of each fermionic species $x$ ($x \in
\{F,\Psi\}$). Ignoring the thermal distribution of bosons, which does not depend
on $\varphi$, the total free energy is
\be
\label{FreeEnergy}
\Phi(\varphi) \;\;=\;\; - \mu^b\,\varphi^2\;\; + \;\;\lambda\,\varphi^4\;\;
+ \sum_{x \in \{F,\Psi\}}
R^x(\varphi) \punc{,}
\ee
where
\be
\label{DefineR}
R^x(\varphi) = -\frac{1}{\beta}\int\frac{d^3 k}{(2\pi)^3}\,\ln{\left(1+e^{-
\beta \xi^x_k}\right)}\punc{.}
\ee

The phase transition to a state with nonzero $\varphi$ occurs when the quadratic
coefficient changes sign, \ie\ when
\be
\Delta \equiv \frac{1}{2}{\left.\frac{d^2 \Phi}{d\varphi^2}\right|}_{\varphi=0}
= 0\punc{.}\ee
Specifically, for negative $\Delta$, nonzero $\varphi$ is energetically favored,
so that the condensed phase is stable. Using (\ref{MixedDispersion}),
(\ref{FreeEnergy}) and (\ref{DefineR}), we find
\be
\label{DiscriminantIntegral}
\Delta = -\mu^b + g^2\int\frac{d^3 k}{(2\pi)^3}\,\frac{n\sub{F}(\xi^f_k)-
n\sub{F}(\xi^\psi_k)}{\xi^f_k-\xi^\psi_k}\punc{,}
\ee
where $n\sub{F}$ is the Fermi-Dirac distribution function. The integral equation
$\Delta = 0$ may be solved numerically.

\subsection{Particle numbers}
\label{MFFixedN}

Since experiments are necessarily performed at fixed particle number, the
expressions for the numbers in terms of the chemical potentials must be found.
Particles of the species $b$, $f$ and $\psi$ are not independently
conserved, so the relevant quantities are $N_f$ and $N_b$, the total numbers of
fermionic and bosonic atoms, respectively (including those contained in molecules).

As mentioned above, it is sufficient within mean-field theory to determine these
numbers to order zero in the coupling. Since the species $F$ and $\Psi$ each
contain one atomic fermion, we have
\be
N_f = \int\frac{d^3 k}{(2\pi)^3}\,\left[n\sub{F}(\xi^\Psi_k) +
n\sub{F}(\xi^F_k)\right]\punc{.}
\ee
The number of bosons is (with $n\sub{B}$ the Bose-Einstein distribution
function)
\bea
N_b = \varphi^2 + \int\frac{d^3 k}{(2\pi)^3}\,\big[n\sub{B}(\xi^b_k) &+&
n\sub{F}(\xi^\Psi_k)\cos^2\theta_k\\
&+&n\sub{F}(\xi^F_k)\sin^2\theta_k\big]\punc{,}\nonumber
\eea
where the first term represents the condensate, the first term in the integrand
is the thermal distribution of the bosons, and $\theta_k$ is the mixing
angle.\footnote{Explicitly, $\theta_k$ is the angle parametrizing the unitary
transformation from the fermions $f$ and $\psi$ in (\ref{HamBefore}) to the
fermions $F$ and $\Psi$ in (\ref{HamAfter}).}

When $\varphi = 0$, such as along the boundary to the phase without
a BEC, the expression for the number of bosons simplifies to
\be
N_b = \int\frac{d^3 k}{(2\pi)^3}\,\left[n\sub{B}(\xi^b_k) +
n\sub{F}(\xi^\psi_k)\right]\punc{.}
\ee
To locate this phase boundary for fixed particle numbers, we must find the values
of $\mu^f$ and $\mu^b$ which give the required numbers and also satisfy $\Delta
= 0$. (Of course, a third parameter must be tuned to its critical value to
satisfy these three conditions simultaneously.)

\begin{figure}
        \resizebox{\columnwidth}{!}{
            \includegraphics{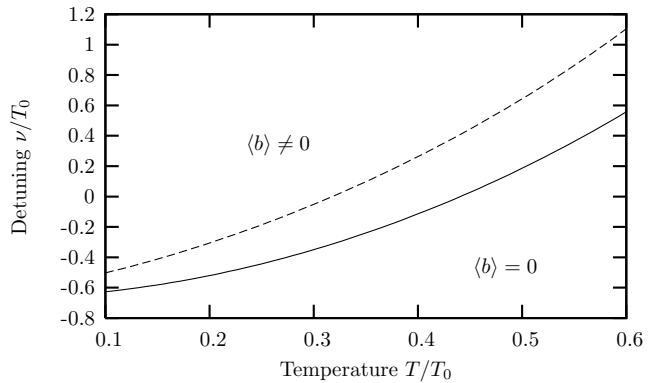}
        }
\caption{\label{Graph0}Phase boundary with detuning $\nu$ and temperature $T$,
for fixed particle numbers $N_f/N_b = 1.11$. The dashed line has
vanishing coupling and has been found with a purely classical analysis. The
solid line has dimensionless coupling
$\gamma^2/T_0 = 2.5\times 10^{-4}$, and has been determined using the mean-field
theory of Section \ref{MFforMu}. For both, the condensed phase is on the
left-hand side (for lower $T$) and labeled by $\Mean{b}\neq 0$.}
\end{figure}
Results from such a procedure are displayed in Fig.~\ref{Graph0}, which shows
the boundary for $N_f/N_b = 1.11$ as a function of the detuning $\nu$
and temperature $T = 1/\beta$. The masses of the atoms are equal, $m^f = m^b$, and
the solid line has dimensionless coupling $\gamma^2/T_0 = 2.5\times 10^{-4}$. For
comparison, the case of vanishing
coupling, treated in Section \ref{Zeroglimit}, is also shown with a dashed line.
Both curves reach the value $T = T_0$, as in the case of free bosons, for
$\nu\rightarrow\infty$, when molecules cannot be formed.

\begin{figure}
        \resizebox{\columnwidth}{!}{
            \includegraphics{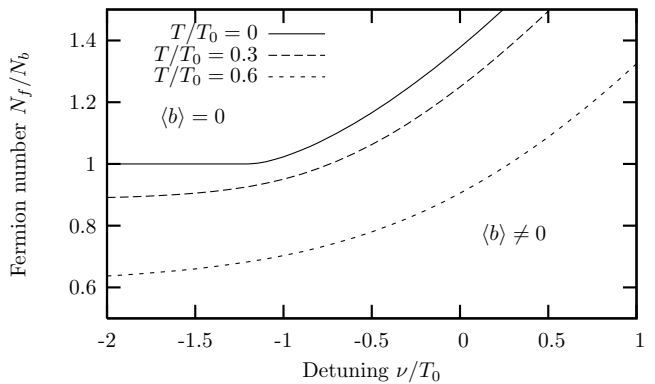}
        }
\caption{\label{Phases}Phase boundary with fermion number $N_f$ and detuning $\nu$,
for three different temperatures. The coupling is $\gamma^2/T_0 = 2.5\times 10^{-4}$
and the masses are equal, $m^f = m^b$. The two phases are labeled as in
Fig.~\ref{Graph0}, with the condensed phase favored for higher detuning, lower
fermion number and lower temperature.}
\end{figure}

In Fig.~\ref{Phases} the same phase boundary is shown on a graph of fermion number
versus detuning, for three different temperature values. The solid line is at zero
temperature, $T = 0$, while the two dashed lines have nonzero temperatures. The
coupling is $\gamma^2/T_0 = 2.5\times 10^{-4}$ and the masses are equal, $m^f = m^b$.
As expected, Bose condensation is favored by lower temperatures, as in the case of
an isolated Bose gas.

It remains to be shown that the system is stable against separation into regions
with different densities. It is shown in the Appendix that it is indeed stable
for a large range of parameter values.

\subsection{Zero-temperature phases}
\label{ZeroT}

At $T = 0$, the Fermi-Dirac distribution function is replaced by a unit step
and all bosons occupy the lowest-energy state. As noted by Yabu {\em
et al.}, \cite{Yabu}
the phase diagram can be further divided into a
region with two Fermi surfaces and a region with a single Fermi surface. (We ignore
the trivial case without any Fermi surfaces, which requires $N_f = 0$.)

Except when the atomic numbers precisely match, $N_f = N_b$, the case of a single
surface can only occur when there is a BEC. In this case, $\varphi$, the
expectation value of $b$, is given by the minimum of the free energy $\Phi$ given
in (\ref{FreeEnergy}), so that we must solve
\be
\label{EqCondensate}
- 2\mu^b\,\varphi\;\; + \;\;4\lambda\,\varphi^3\;\;
+ \sum_{x \in \{F,\Psi\}}
\frac{dR^x}{d\varphi} \;=\; 0
\ee
(excluding the root $\varphi = 0$).

Following the choice that $\xi^F_k\ge\xi^\Psi_k$ in (\ref{MixedDispersion}), the
second Fermi surface disappears when $\xi^F_{k=0} = 0$, making the Fermi
wavenumber for $F$ fermions vanish. For this to be the case, we require $\mu^f > 0$,
$\mu^\psi > \nu$ and
\be
g\varphi = \sqrt{\mu^f(\mu^\psi - \nu)}\punc{,}
\ee
which should be solved simultaneously with (\ref{EqCondensate}).

\begin{figure}
    \resizebox{\columnwidth}{!}{
        \includegraphics{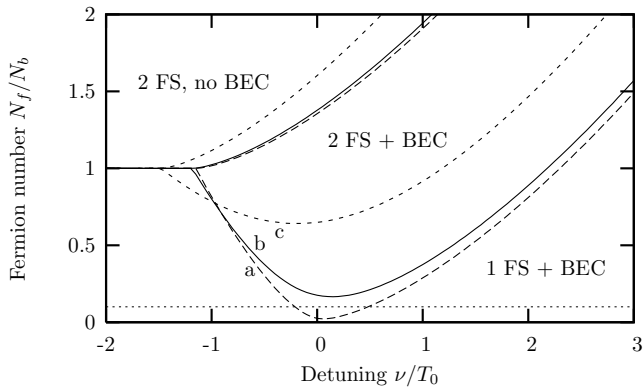}
    }
\caption{\label{ZeroTphases}The phase diagram at $T = 0$ with dimensionless
couplings (a) $\gamma^2/T_0 = 10^{-6}$, (b) $\gamma^2/T_0 = 2.5\times 10^{-4}$ and
(c) $\gamma^2/T_0 = 2.0\times 10^{-2}$. The atomic masses have been taken to be equal,
$m^f = m^b$ and the coupling between bosons is given by
$\lambda^2 (m^b)^3 T_0 = 2\times 10^{-3}$.
The three distinct phases have, respectively, no Bose-Einstein
condensate and two Fermi surfaces (labeled `2 FS, no BEC'), a condensate and two
Fermi surfaces (`2 FS + BEC'), and a condensate and a single Fermi surface
(`1 FS + BEC'). The dotted line indicates the fermion number at which
Fig.~\ref{ZeroTmass} is plotted.}
\end{figure}
These expressions, along with the results in Section \ref{MFFixedN} for the particle
numbers, allow the complete zero-temperature phase diagram to be plotted. In
Fig.~\ref{ZeroTphases}, the phase boundaries are shown on a graph of fermion number
against detuning, for equal atomic masses $m^f = m^b$.
The three sets of boundaries have couplings (a) $\gamma^2/T_0 = 10^{-6}$, (b)
$\gamma^2/T_0 = 2.5\times 10^{-4}$ and (c) $\gamma^2/T_0 = 2.0\times 10^{-2}$.
(Since the dimensionless coupling depends on the fourth
power of the coupling $g$ appearing in the Hamiltonian, these large changes in
$\gamma^2/T_0$ in fact correspond to changes in $g$ of only factors of $4$ and $3$
respectively. All of these coupling values are within the narrow resonance regime.)
Throughout, we take $\lambda^2 (m^b)^3 T_0 = 2\times 10^{-3}$.

The boundaries divide the diagram into three regions, depending on the presence of a
condensate and the number of Fermi surfaces. In the region labeled `2 FS, no BEC',
the discriminant $\Delta$ is positive, so there is no BEC and
two Fermi surfaces. In the region labeled `2 FS + BEC', $\Delta$ is negative and
there is a condensate, as well as two Fermi surfaces. The lowermost region of the
diagram, labeled `1 FS + BEC', has a condensate and only a single Fermi surface.

In the limit of vanishing coupling (as in Ref.~\onlinecite{Yabu} and Section
\ref{Zeroglimit}), the boundary between the regions
with one and two Fermi surfaces extends down to the line $N_f = 0$. The region with
a single Fermi surface is then divided into two, with the left-hand side having
a Fermi surface of molecules and the right-hand side a Fermi surface of atoms.
Including the quantum-mechanical effects, these two regions are no longer distinct,
with the single Fermi surface crossing over from having a molecular character on
one side (lower $\nu$) to having an atomic character on the other (higher $\nu$).

\begin{figure}
    \resizebox{\columnwidth}{!}{
        \includegraphics{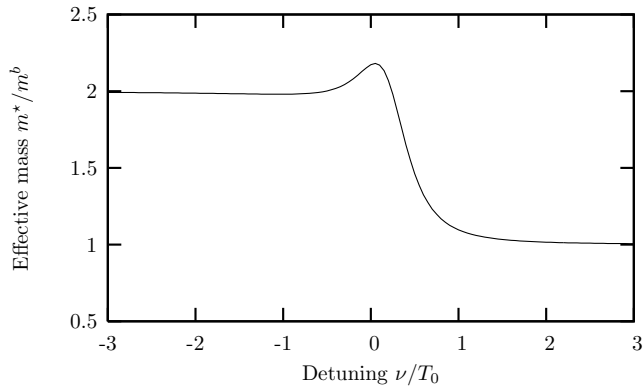}
    }
\caption{\label{ZeroTmass}The effective mass $m^\star$ at the Fermi surface,
with fermion number $N_f = 0.1 N_b$, coupling $\gamma^2/T_0 = 2.5\times 10^{-4}$
and equal atomic masses $m^f = m^b$. As can be seen from the dotted line in
Fig.~\ref{ZeroTphases}, these parameters give a phase with a single
Fermi surface. This surface changes from having a molecular character, with
$m^\star\simeq m^\psi$, to having an atomic character, $m^\star\simeq m^f$.}
\end{figure}

This crossover is illustrated in Fig.~\ref{ZeroTmass}, where the effective mass
$m^\star$ at the Fermi surface is plotted. The fermion number is set at
$N_f = 0.1N_b$ and the coupling is $\gamma^2/T_0 = 2.5\times 10^{-4}$, so that the
system is within the phase with a single Fermi surface (of $\Psi$ fermions). The
effective mass is defined as
\be
m^\star = {\left({\left.\frac{d^2\xi^\Psi_k}{dk^2}\right|}_{k_0^\Psi}\right)}^{-1}\punc{.}
\ee
For $\nu\ll 0$, the Fermi surface has an essentially molecular character and
$m^\star\simeq m^\psi$, while for $\nu\gg 0$, it is atom-like, with
$m^\star\simeq m^f$.

\begin{figure}
    \resizebox{\columnwidth}{!}{
        \includegraphics{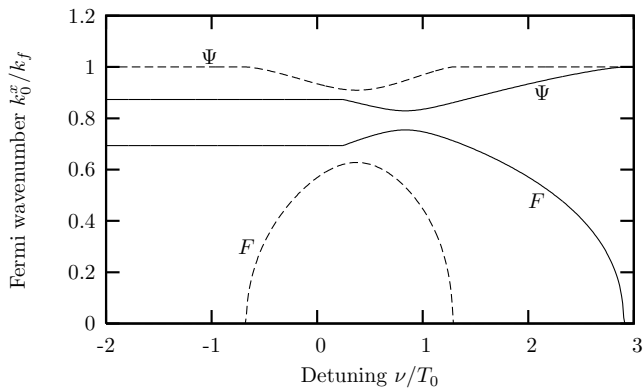}
    }
\caption{\label{ZeroTk}The Fermi wavenumbers for the two mixed species of fermions,
$\Psi$ and $F$, with coupling $\gamma^2/T_0 = 2.5\times 10^{-4}$ and equal atomic
masses $m^f = m^b$. The solid lines have fermion number $N_f = \frac{3}{2} N_b$,
while the dashed lines have $N_f = \frac{1}{2} N_b$. As can be seen in
Fig.~\ref{ZeroTphases}, the solid line goes between all three phases (at
$\nu/T_0\simeq 0.25$ and $\nu/T_0\simeq 2.9$), while the dashed line goes from
the phase with a single Fermi surface to that having two and back again (at
$\nu/T_0\simeq -0.65$ and $\nu/T_0\simeq 1.3$). The wavenumbers are measured
in units of $k_f$, the Fermi wavenumber for free fermions with number $N_f$.}
\end{figure}

In Fig.~\ref{ZeroTk}, the Fermi wavenumbers of the two fermionic species are plotted,
for coupling $\gamma^2/T_0 = 2.5\times 10^{-4}$ and two different fermion numbers,
$N_f = \frac{3}{2} N_b$ (solid lines) and $N_f = \frac{1}{2} N_b$ (dashed lines).
In both the phase without a condensate (solid lines for $\nu/T_0 < 0.25$) and the
phase with a single Fermi surface (solid lines for $\nu/T_0 > 2.9$, dashed lines
for $\nu/T_0 < -0.65$ and $\nu/T_0 > 1.3$) the wavenumbers are constant, due to the
fixed particle numbers. Only in the phase with two Fermi surfaces and a BEC
do the Fermi wavevectors change with detuning.
(At the fermion number used in Fig.~\ref{ZeroTmass}, the system stays in the phase
with a single Fermi surface throughout and $k_0^\Psi = k_f$, $k_0^F = 0$ for all
detunings.)

We now turn our attention to the line dividing the phases 2 FS, no BEC and
1 FS + BEC in Fig.~\ref{ZeroTphases}. This boundary is horizontal and starts
at the point where the three phases meet; in Section \ref{Multi}, we will prove
that this is at exactly $N_b = N_f$. At this transition, two changes
occur: both the second Fermi surface vanishes and the BEC
appears, as the line is crossed from above. Physically, this results from the fact
that molecules are highly energetically favored in this region, so that as many
molecules as possible are formed, and the residual atoms form their ground state.
For $N_f > N_b$, these atoms are fermionic and form a Fermi surface, while for
$N_f < N_b$, they are bosonic and form a condensate. Precisely at $N_f = N_b$,
there are no residual atoms, so that there is no condensate and only a molecular
Fermi surface.

\begin{figure}
    \resizebox{\columnwidth}{!}{
        \includegraphics{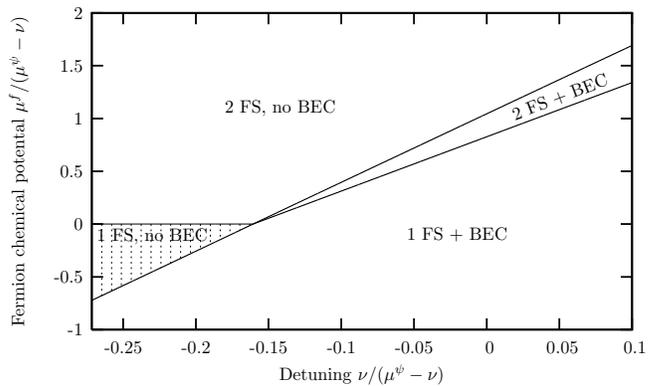}
    }
\caption{\label{ZeroTmu}The phase diagram with the fermion
chemical potential $\mu^f$ plotted on the vertical axis and the
detuning $\nu$ on the horizontal axis. Both have been scaled to
$\mu^\psi - \nu$, which is held fixed throughout the plot. The
boundary between 2 FS, no BEC and 1 FS + BEC in
Fig.~\ref{ZeroTphases} expands into a new phase, labeled `1 FS, no
BEC', within which there are only molecules, whose density is
constant (both $N_f$ and $N_b$ remain fixed in the shaded region).
The atomic masses are equal, $m^f = m^b$, and the couplings are
$\gamma^2/T_0 = 2.5\times 10^{-4}$ and $\lambda^2 (m^b)^3 T_0 =
2\times 10^{-3}$. (Since the boson density is not fixed in this
plot, the value of $T_0$ appropriate to the phase 1 FS, no BEC has
been used to define the dimensionless couplings.)}
\end{figure}

This situation is illustrated by Fig.~\ref{ZeroTmu}, which shows
the same phase diagram as Fig.~\ref{ZeroTphases}, but with the
chemical potential for the fermionic atoms, $\mu^f$, on the
vertical axis. Throughout the plot $\mu^\psi - \nu$, and hence the
Fermi wavenumber for the molecules, $k_0^\psi$, is held fixed. In
the region where $\mu^f > 0$, the essential features are
unchanged, with the same three phases as shown in
Fig.~\ref{ZeroTphases}. The boundary between the phases 2 FS, no
BEC and 1 FS + BEC, however, is seen to extend into an entire
phase, labeled `1 FS, no BEC'. In this region, there is no
condensate and $\mu^f$ is negative, so that there is only one
Fermi surface, of molecules. This entire phase therefore has $N_f
= N_b$ and collapses onto a single line in Fig.~\ref{ZeroTphases}.
Moreover, because $k_0^\psi$ is constant, $N_f$ and $N_b$ are both
fixed in this phase. The situation in the shaded region of
Fig.~\ref{ZeroTmu} resembles that in the Mott insulator lobes in
the phase diagram of the boson Hubbard model (see
Ref.~\onlinecite{fwgf} and Chapter 10 of Ref.~\onlinecite{book});
at fixed $\mu^\psi$, the density of particles is insensitive to
the variation in the chemical potential $\mu^f$.

\section{Luttinger's theorem}
\label{luttinger}

All the ground states in our phase diagram in
Fig.~\ref{ZeroTphases} contain Fermi surfaces. In
Fig.~\ref{ZeroTk} we presented the evolution of the Fermi
wavevectors of these Fermi surfaces in our mean-field calculation.
In the present section we will discuss general constraints that
must be satisfied by these Fermi wavevectors which are valid to
all orders in the interactions. (Throughout this section, we shall
be concerned only with $T=0$.)

We will base our arguments upon the existence of the Luttinger-Ward
functional \cite{luttingerward} $\LW$, satisfying
\be
\label{EqDefineLW}
\Sigma = {\left.\frac{\delta \LW[G']}{\delta G'}\right|}_{G'=G}\punc{,}
\ee
where $G'$ is a dummy variable, $G$ is the actual full (thermal) Green function
and $\Sigma$ is the full self-energy. (Throughout this section, we will
mostly be concerned with the full Green functions, which we shall denote
with the symbol $G$. When we make reference to the free Green function,
this will be denoted $G_0$.)

Following Ref.~\onlinecite{potthoff},
it possible to construct the Luttinger-Ward functional non-perturbatively,
starting from the
partition function $Z$ of the system. It can be shown straightforwardly
that, treating the Green function as a matrix in its momentum (and frequency)
indices, any unitary transformation of the free Green function,
$G_0 \rightarrow UG_0 U^{-1}$, that leaves $Z$ invariant also
leaves $\LW$ invariant.

A standard proof of Luttinger's theorem \cite{agd}
for a system of interacting fermions makes use of the invariance of $Z$ under
a shift in the frequency appearing in the free propagator,
$\omega\rightarrow \omega + \alpha$. In our case $\LW$ is a functional
of the three Green functions, one for each species, and $Z$ is invariant under
a simultaneous shift in two of the three frequencies, \ie
\bea
&\LW[G^\psi(i\nu_1),G^f(i\nu_2),G^b(i\omega)]&\nonumber\\
=&\LW[G^\psi(i\nu_1),G^f(i\nu_2-i\alpha),G^b(i\omega+i\alpha)]&\nonumber\\
=&\LW[G^\psi(i\nu_1+i\beta),G^f(i\nu_2),G^b(i\omega+i\beta)]&
\label{EqInvariance}
\eea
for any $\alpha$ and $\beta$.

To proceed further, it is useful to set $\mu^b = \mu^\psi - \mu^f$ and
consider derivatives of the grand energy with respect to $\mu^f$
and $\mu^\psi$. The derivative with respect to $\mu^f$ yields
\be
\Mean{f^\dagger f} - \Mean{b^\dagger b} = N_f - N_b\punc{.}
\label{const1}
\ee
Each term on the left-hand side can be rewritten in terms of the full
Green functions, giving
\be
\label{EqNumbersFromGreenFunc}
N_f - N_b = -\int\frac{d^3k\,d\omega}{(2\pi)^4}e^{i\omega 0^+}
\left[G^f_k(i\omega) + G^b_k(i\omega)\right]\punc{.}
\ee
(The change of sign of the $f$ term results from the anticommutation
of fermion operators.)

From now on the manipulations are standard.\cite{agd} We make use of the identity
\be
G(i\omega) = iG(i\omega)\parder{}{\omega}\Sigma(i\omega)
- \parder{}{\omega}\ln{G(i\omega)}\punc{,}
\ee
which results from the Dyson equation. The first equation of
(\ref{EqInvariance}) gives, together with (\ref{EqDefineLW}),
\be
\int\frac{d^3k\,d\omega}{(2\pi)^4}\left[
\Sigma^b_k(i\omega)\parder{}{\omega}G^b_k(i\omega) +
\Sigma^f_k(i\omega)\parder{}{\omega}G^f_k(i\omega)
\right] = 0\punc{.}
\ee
Combining these two with (\ref{EqNumbersFromGreenFunc}) and integrating by
parts gives
\be
N_f - N_b = i\int\frac{d^3k\,d\omega}{(2\pi)^4}e^{i\omega 0^+}
\parder{}{\omega}\left[\ln G^f_k(i\omega) + \ln G^b_k(i\omega)\right]
\ee
(with the boundary terms vanishing because $G(i\omega)\sim 1/|\omega|$ for
$|\omega|\rightarrow\infty$).

The integral over $\omega$ can be treated as a contour integration and
closed above, due to the factor $e^{i\omega 0^+}$. Changing the integration
variable to $z = i\omega$ replaces this by an integral surrounding the left
half-plane. Since both of the full Green functions $G^{b,f}_k(z)$ have all
their non-analyticities and zeroes on the line of real $z$, we can write this as
\begin{widetext}
\be
N_f - N_b = i\int\frac{d^3k}{(2\pi)^3}\int_{-\infty}^{0}\frac{dz}{2\pi}
\parder{}{z}\left[\ln G^f_k(z+i0^+) + \ln G^b_k(z+i0^+)
- \ln G^f_k(z+i0^-) - \ln G^b_k(z+i0^-)\right]\punc{.}
\ee
\end{widetext}
The integral of $z$ can be performed trivially to give
\be
N_f - N_b = i\int\frac{d^3k}{(2\pi)^4}
\left[\ln \frac{G^f_k(i0^+)}{G^f_k(i0^-)}
+\ln \frac{G^b_k(i0^+)}{G^b_k(i0^-)}
\right]\punc{.}
\ee

Using the analyticity properties of the Green functions, this gives
\be
\label{EqStepFunctions}
N_f - N_b = \int\frac{d^3k}{(2\pi)^3}\left[\Theta(-\xi^f_k+\Sigma'^f_k)
+\Theta(-\xi^b_k+\Sigma'^b_k)\right]\punc{,}
\ee
where $\Theta$ is the unit step function and $\Sigma'$ is the real part of the
self-energy evaluated for $\omega = 0$.

First, we consider the phase with no BEC. Here there are
necessarily two Fermi surfaces, and, as we will now show, the
volumes of the two Fermi surfaces are separately constrained, independently
of the interactions.

By definition, the absence of a BEC requires that there be no bosonic
quasiparticle excitations at or above the chemical potential, so that the second
term in the brackets in (\ref{EqStepFunctions}) vanishes. (Note that this does
not imply that $\Mean{b^\dagger b} = 0$, which is not the case beyond mean-field
order.) This leaves the statement of Luttinger's theorem for this case:
\be
\label{EqLutt1}
N_f - N_b = \int\frac{d^3k}{(2\pi)^3}\Theta(-\xi^f_k+\Sigma'^f_k)\punc{.}
\ee
The right-hand side of this expression is interpreted as the (reciprocal-space)
volume of the atomic Fermi surface.

A similar result follows by taking the derivative of the grand energy
with respect to $\mu^\psi$, which gives
\be
\Mean{\psi^\dagger \psi} + \Mean{b^\dagger b} = N_b\punc{.}
\label{const2}
\ee
Going through the same manipulations as above leads to
\be
\label{EqStepFunctions2}
N_b = \int\frac{d^3k}{(2\pi)^3}\left[\Theta(-\xi^\psi_k+\Sigma'^\psi_k)
-\Theta(-\xi^b_k+\Sigma'^b_k)\right]\punc{,}
\ee
corresponding to (\ref{EqStepFunctions}). Since we are in the phase with no BEC,
this gives
\be
\label{EqLutt2}
N_b = \int\frac{d^3k}{(2\pi)^3}\Theta(-\xi^\psi_k+\Sigma'^\psi_k)\punc{.}
\ee

We have therefore proved that there are two statements of Luttinger's theorem
in the phase with two Fermi surfaces and no BEC. One, (\ref{EqLutt1}), states that
the volume of the atomic Fermi surface is fixed by the difference in the numbers
of atomic fermions and bosons, while the other, (\ref{EqLutt2}), states that the
volume of the molecular Fermi surface is fixed by the total number of bosonic atoms.

Now let us extend our considerations to the phases with a BEC. In
Fig.~\ref{ZeroTphases} we observe that these phases may have
either one or two Fermi surfaces. Here we show that Luttinger's
theorem only demands that the total volume enclosed within {\em
both} Fermi surfaces is determined by $N_f$; the volumes of the two
Fermi surfaces (if present) are not constrained separately.

In the presence of a BEC, it is no longer the case that the second
term in the brackets vanishes in (\ref{EqStepFunctions}) and (\ref{EqStepFunctions2}).
Instead, if we add these two equations, we arrive at
\be
\label{EqLutt3}
N_f = \int\frac{d^3k}{(2\pi)^3}\Theta(-\xi^f_k+\Sigma'^f_k)
+ \int\frac{d^3k}{(2\pi)^3}\Theta(-\xi^\psi_k+\Sigma'^\psi_k)\punc{.}
\ee
The two terms in this expression give the volumes of the two Fermi surfaces.
We see that their sum is constrained to equal the number of fermionic atoms.

\subsection{Multicritical point}
\label{Multi}

A simple application of our statements of Luttinger's theorem can
be used to show that the multicritical point, where the three
phases meet in Fig.~\ref{ZeroTphases} (and where four phases meet
in Fig.~\ref{ZeroTmu}), occurs at precisely $N_b = N_f$.

Firstly, according to (\ref{EqLutt1}), the volume of the atomic
Fermi surface is given by $N_f - N_b$, as long as there is no BEC.
This is therefore the case on the line dividing the phases with
and without condensates, since the condensate vanishes as this
line is approached from below. Secondly, the line that divides the
regions with one and two Fermi surfaces is the point where the
atomic Fermi surface vanishes. Where the two lines meet, we see
both that (\ref{EqLutt1}) is satisfied and that its right-hand
side vanishes. We therefore have $N_b = N_f$.

\section{Quantum phase transitions}
\label{EffActBoson}

We now present an alternative analysis using the language of field
theory. In Section \ref{MF2}, we reproduce the result that
(\ref{DiscriminantIntegral}) determines the presence of the
condensate. Then, in Section \ref{SecBosonPropagator}, we
determine the boson propagator near the BEC depletion transition.
In Section \ref{critical}, we describe the critical field theories
for the various transitions.

The dimensionless Euclidean action corresponding to the Hamiltonian
(\ref{Hamiltonian}) is
\bea
\Act &= &\frac{1}{\beta}\sum_q \bar{f}_q \Xi^f_q f_q + \frac{1}{\beta}\sum_p
\bar{b}_p \Xi^b_p b_p + \frac{1}{\beta}\sum_q \bar{\psi}_q
\Xi^\psi_q\psi_q\nonumber\\
&& -\: \frac{g}{\beta^2}\sum_{p,q} \left(\bar{\psi}_q f_{q-p} b_p +
\bar{b}_p\bar{f}_{q-p}\psi_p\right)\punc{.}
\label{eqAction}
\eea
The symbol $p$ stands for $k$ and $\omega_n$, and likewise $q$ for $\ell$ and
$\nu_m$, where the Matsubara frequencies $\omega_n$ ($\nu_m$) are even (odd).
The summations over $p$ ($q$) represent sums over $\omega_n$ ($\nu_m$) and
integrals over the momentum $k$ ($\ell$). We have also defined
\be
\Xi_p = {\left(G_p\right)}^{-1} = -i\omega_n + \xi_k\punc{,}
\ee
the inverse of the free Green function, $G_p$, and similarly $\Xi_q$. (In this
section and the following, we will use the symbol $G$ to denote the free Green
function, contrary to the notation of Section \ref{luttinger}.)

We have omitted from the action the coupling term between pairs of bosons,
since we will be interested in the region near the phase transition, where
this term is not important.

Integrating out both of the fermions and expanding the resulting coupling term
to quadratic order in $b$ and $\bar{b}$, we find that the effective action for
the bosons is
\be
\label{QuadAction}
\Act^{(2)}\sub{eff}[b,\bar{b}] = \frac{1}{\beta}\sum_p\bar{b}_p\Xi^b_p b_p +
\frac{g^2}{\beta^2}\sum_{p,q}G^f_q G^\psi_{q+p} \bar{b}_{p} b_{p}\punc{.}
\ee

\subsection{Mean-field approximation}
\label{MF2}

By replacing $b$ by a real constant $\varphi$, we should arrive at the results
of Section \ref{MFforMu}. In this approximation, we have
\be
\Act^{(2)}\sub{eff}[b,\bar{b}] = -\mu^b \varphi^2 +
\frac{g^2}{\beta}\varphi^2\sum_{q}G^f_q G^\psi_q\punc{,}
\ee
so that the coefficient is
\be
\label{DeltaInFieldTheory}
\Delta = -\mu^b + g^2\int\frac{d^3
\ell}{(2\pi)^3}\frac{1}{\beta}\sum_{\nu_m}G^f_\ell(i\nu_m)\,
G^\psi_\ell(i\nu_m)\label{MeanFieldProp}\punc{.}
\ee
The phase transition will occur when the coefficient $\Delta$ vanishes.

We can represent (\ref{DeltaInFieldTheory}) by
\be
\parbox{35mm}{
\begin{picture}(35,20)
\put(0,0){\includegraphics{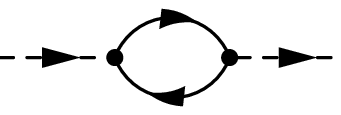}}
\input{fmfPaper.t1}
\end{picture}
}\punc{,}
\label{FirstGraph}
\ee
which appears as a self-energy diagram in the boson propagator, drawn as a
dashed line. (The two solid lines represent fermion propagators.)

The Matsubara sum can be performed by replacing it by a contour integration,
giving
\be
\Delta = -\mu^b + g^2\int\frac{d^3 \ell}{(2\pi)^3}\,\frac{n\sub{F}(\xi^f_\ell)-
n\sub{F}(\xi^\psi_\ell)}{\xi^f_\ell-\xi^\psi_\ell}\punc{,}
\ee
in agreement with (\ref{DiscriminantIntegral}).

\subsection{Boson propagator}
\label{SecBosonPropagator}

By retaining the frequency dependence of the boson field, but again keeping
only terms quadratic in $b$ and $\bar{b}$, we can determine the form of the
long-wavelength, low-frequency excitations.

The effective boson propagator is, from (\ref{QuadAction}), the reciprocal of
\bea
\Xit^b_k(i\omega_n)&\equiv&\Xit^b_p\;\equiv\;\Xi^b_p +
\frac{g^2}{\beta}\sum_q G^f_{q} G^\psi_{q+p}\label{EffBoseDisp}\\
&=&-i\omega_n+\xi^b_k + g^2\int\frac{d^3
\ell}{(2\pi)^3}\times\nonumber\\
&&\frac{1}{\beta}\sum_{\nu_m}\frac{1}{-
i\nu_m+\xi^f_{\ell}}\frac{1}{-
i(\nu_m+\omega_n)+\xi^\psi_{\ell+k}}\nonumber\punc{,}
\eea
which replaces (\ref{MeanFieldProp}). For $k=0$, this gives
\be
\Xit^b_0(i\omega_n)=-i\omega_n-\mu^b+g^2\int\frac{d^3
\ell}{(2\pi)^3}\,\frac{n\sub{F}(\xi^f_\ell)-n\sub{F}(\xi^\psi_\ell)}{\xi^f_\ell-
\xi^\psi_\ell+i\omega_n}
\punc{,}
\ee
where the result
\be
n\sub{F}(a - i\omega_n) = n\sub{F}(a)\punc{,}
\ee
for $\omega_n$ a boson Matsubara frequency, has been used. For small $\omega_n$,
we can expand to give
\bea
\Xit^b_0(i\omega_n) &\simeq& \Delta - i\omega_n\left[ 1 - g^2\int\frac{d^3
\ell}{(2\pi)^3}\frac{n\sub{F}(\xi^f_\ell)-n\sub{F}(\xi^\psi_\ell)}{{(\xi^f_\ell
- \xi^\psi_\ell)}^2} \right]\nonumber\\
&&-\;\omega_n^2 g^2\int\frac{d^3 \ell}{(2\pi)^3}\frac{n\sub{F}(\xi^f_\ell)-
n\sub{F}(\xi^\psi_\ell)}{{(\xi^f_\ell - \xi^\psi_\ell)}^3}
\punc{.}
\eea
(Note that, as required, the coefficient of $\omega_n^2$ is in fact positive.)

The effective boson propagator (for $k=0$) is then
\be
\label{Gtildeb}
\Gt^b_0(i\omega_n)=\frac{\mathcal{Z}}{-i\omega_n + \xit^b_0(\omega_n)}\punc{,}
\ee
with
\be
\mathcal{Z} = {\left[ 1 - g^2\int\frac{d^3
\ell}{(2\pi)^3}\frac{n\sub{F}(\xi^f_\ell)-n\sub{F}(\xi^\psi_\ell)}{{(\xi^f_\ell
- \xi^\psi_\ell)}^2}\right]}^{-1}\punc{,}
\ee
and
\be
\xit^b_0(\omega_n)=\mathcal{Z}\left[\Delta - \omega_n^2 g^2\int\frac{d^3
\ell}{(2\pi)^3}\frac{n\sub{F}(\xi^f_\ell)-n\sub{F}(\xi^\psi_\ell)}{{(\xi^f_\ell
- \xi^\psi_\ell)}^3}\right]\punc{.}
\ee

The integrals in the expressions for both $\mathcal{Z}$ and
$\xit^b_0(\omega_n)$ diverge at zero temperature if $N_f = 2 N_b$
so that the two Fermi wavenumbers coincide. For any other
parameters the integrals are finite, and the effective propagator
has the form (\ref{Gtildeb}). As we discuss in the following
subsection, this distinction leads to different field theories for
the BEC depletion quantum transition for these cases.

\subsection{Critical field theories}
\label{critical}

First, we consider the BEC depletion quantum transition. This is
the transition between the 2 FS + BEC phase and the 2 FS, no BEC
phase in Fig.~\ref{ZeroTphases}. The same theory also applies to
the transition between the 1 FS + BEC phase and the 1 FS, no BEC
phase in Fig.~\ref{ZeroTmu}. The low momentum modes of the $b$
boson field clearly constitute an order parameter for this
transition. The effective action for the renormalized $b$ field
near the critical point can be derived by integrating out the
fermionic degrees of freedom, as already outlined in
Section~\ref{SecBosonPropagator}.

For $N_f \neq 2 N_b$, the integration of the fermionic excitations
is entirely free of infrared singularities: the differences in the
two Fermi wavevectors implies that there are no low momentum
fermionic particle-hole excitations at low energies. The resulting
action for $b$ contains only terms which are analytic in frequency
and momentum, and has the following familiar form:
\begin{eqnarray}
\mathcal{S}_c [b] &=& \int d^3 r \int d \tau \Biggl[ b^\dagger
\frac{\partial b}{\partial \tau} - \frac{1}{2\tilde{m}^b} b^\dagger \nabla^2
b + s |b|^2 \nonumber \\ &~&~~~~~~~~~~~~~~~~~~~~~~~~~+ \frac{u}{2}
|b|^4 \Biggr] \label{scb}
\end{eqnarray}
Note that the $b$ field has been rescaled by a factor $\sqrt{\mathcal{Z}}$
from the $b$ field in Section~\ref{SecBosonPropagator} and that its
mass has been replaced by the renormalized mass $\tilde{m}^b$.
The action $\mathcal{S}_c
[b]$ describes a quantum phase transition with dynamic exponent
$z=2$ driven by tuning the coupling $s$. This transition has been
discussed previously in Ref.~\onlinecite{fwgf} and in Chapter 11
of Ref.~\onlinecite{book}. The upper critical dimension is $d=2$,
above which the quartic coupling $u$ is formally irrelevant.
Nevertheless, the coupling $u$ is important for the $T>0$
crossovers in the vicinity of the quantum critical point: these
are as presented in Ref.~\onlinecite{book}.

For $N_f = 2N_b$, there are low energy fermionic particle-hole
excitations at zero momentum, and so the above procedure has to be
reconsidered. Now there are non-analytic terms in the effective action
for $b$, but these have a structure similar to that found by Hertz
\cite{hertz} for the onset of ferromagnetism in a Fermi liquid.
Evaluating (\ref{EffBoseDisp}) for this case following Hertz, we now
find the effective action
\begin{eqnarray}
\mathcal{S}_c [b] &=& \int d^3 k \int d \omega |b (k, \omega)|^2
\left[ k^2 + c \frac{|\omega|}{k} \right] \nonumber \\
&+& \int d^3 r \int d \tau \Biggl[ s |b|^2 + \frac{u}{2} |b|^4
\Biggr]. \label{scHertz}
\end{eqnarray}
The bare $-i \omega$ term in the boson propagator is not included
above because it is less relevant than the non-analytic term
generated from the Fermi surface excitations. The critical
properties of the $z=3$ critical theory in Eq.~(\ref{scHertz})
have been described earlier by Hertz, and the $T>0$ crossovers by
Millis \cite{millis} (see also Chapter 12 of
Ref.~\onlinecite{book}).

Next, we consider the critical theory of the 2 FS + BEC to 1 FS +
BEC transition in Fig.~\ref{ZeroTphases}. The same theory also
applies to the 2 FS, no BEC to 1 FS, no BEC transition in
Fig.~\ref{ZeroTmu}. Here, a Fermi surface disappears as its Fermi
wavevector vanishes. The critical theory is then the $z=2$ dilute
Fermi gas theory already discussed in Chapter 11 of
Ref.~\onlinecite{book}. All interactions are irrelevant for the
critical properties, and the quantum-critical crossovers are
merely those of a free Fermi gas.

Finally, consider the multi-critical point, noted in
Section~\ref{Multi}, where all phases in Fig.~\ref{ZeroTphases}
and Fig.~\ref{ZeroTmu} meet. Here, both the $b$ bosons and the $f$
fermions are critical. The critical theory is merely the direct
sum of the $z=2$ dilute Bose and Fermi theories mentioned above.
All interactions between the critical $f$ and $b$ modes are
formally irrelevant in three spatial dimensions.

\section{Gaussian corrections}
\label{GaussianCorrections}

In order to test the validity of the approximations made, we shall calculate two
different corrections to the mean-field results of the preceding sections.
First, in this section, we find the corrections to the grand free energy $\Phi$
to order $g^2$. These will result in corrections to the expressions found in
Section \ref{MFFixedN} relating the chemical potentials to the particle numbers.
Subsequently, in Section \ref{LowDensityApproximation}, we will find a new
expression for $\Delta$ by replacing the mean-field theory with a
low-density approximation.

In the remainder of this section, we will use $\Xit^b$ from (\ref{EffBoseDisp})
to determine the corrections to the grand free energy $\Phi$. We will show that
these are negligible, provided that the dimensionless coupling $\gamma^2/T_0$
is sufficiently small. We therefore require a narrow Feshbach resonance for the
results to be quantitatively accurate.

By integrating the
effective action over the boson field, we arrive at an expression for the
partition function including Gaussian corrections,
\be
Z^{(2)} = \frac{(\det \Xi^f) (\det \Xi^\psi)}{(\det \Xit^b)}\punc{.}
\label{StageinF2}
\ee

The grand free energy is then given by
\be
\Phi = \Phi_0^f + \Phi_0^\psi + \frac{1}{\beta}\sum_p \ln \left(\Xi^b_p +
\frac{g^2}{\beta}\sum_q G^f_q G^\psi_{q+p}\right)\punc{,}
\ee
using (\ref{EffBoseDisp}), where
\be
\Phi^x_0 = \pm\frac{1}{\beta}\sum_q \ln \Xi^x_q
\ee
is the grand free energy for the species $x$ in the absence of coupling. [The
plus (minus) sign applies to bosons (fermions).]

Taking a factor of $\Xi^b_p$ out of the logarithm gives $\Phi^b_0$, so that the
correction to $\Phi$ is
\be
\Phi - \Phi_0 = \frac{1}{\beta}\sum_p \ln \left(1 + \frac{g^2}{\beta}\sum_q
G^f_q G^\psi_{q+p} G^b_p\right)\punc{.}
\ee
This is the full expression for the Gaussian corrections; to estimate the size
of these corrections, we will calculate the result to order $g^2$. Dropping terms
of higher order gives a correction to the free energy $\Phi$ of
\be
\Delta\Phi = \frac{g^2}{\beta^2}\sum_{p,q}G^f_q G^\psi_{q+p} G^b_p\punc{,}
\ee
which can be represented diagrammatically as
\be
\Delta \Phi \quad = \quad
\parbox{25mm}{
\begin{picture}(25,25)
\put(0,0){\includegraphics{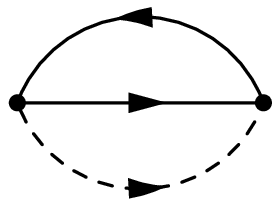}}
\input{fmfPaper.t2}
\end{picture}
}\qquad\punc{.}
\ee

\begin{widetext}
Reinstating explicit momentum integrals and frequency sums, we have
\be
\Delta\Phi=\frac{g^2}{\beta^2}\sum_{\omega_n,\nu_m}\int\frac{d^3
k}{(2\pi)^3}\frac{d^3 \ell}{(2\pi)^3}\:G^b(\omega_n,k)\,G^f(\nu_m,\ell)
\,G^\psi(\nu_m+\omega_n,\ell+k)
\ee
Both Matsubara sums can be performed using contour integration, to give
\be
\Delta \Phi = g^2\,\int\frac{d^3 k}{(2\pi)^3} \int\frac{d^3 \ell}{(2\pi)^3}
\frac{\left[n\sub{F}(\xi^f_{\ell-k})-
n\sub{F}(\xi^\psi_\ell)\right]\left[n\sub{B}(\xi^b_k)-n\sub{B}(\xi^\psi_{\ell} -
\xi^f_{\ell-k})\right]}{\xi^b_k+\xi^f_{\ell-k}-\xi^\psi_{\ell}}\punc{,}
\ee
\end{widetext}
after a change of variables, $\ell\rightarrow\ell-k$.

\subsection{Renormalization of the detuning}
\label{Renormalize}

As it stands, the integral over $k$ is in fact divergent. As $|k|$ tends to
infinity (with $|\ell|$ finite), the second Bose factor, $n\sub{B}(\xi^\psi_\ell
- \xi^f_{\ell-k})$, tends to $-1$. In the first bracket,
$n\sub{F}(\xi^\psi_\ell)$ remains finite, so the integrand tends to $\sim 1/k^2$
and the integral over $k$ is linearly divergent.

This divergence can be understood by considering the self-energy diagram
\be
\label{BubbleDiagram1}
\parbox{35mm}{
\begin{picture}(35,20)
\put(0,0){\includegraphics{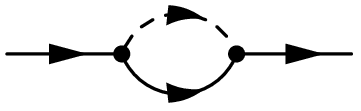}}
\input{fmfPaper.t3}
\end{picture}
}
\ee
which gives a correction to the detuning $\nu$ linear in the cut-off momentum,
\be
\nu = \nu_0 - g^2 \int\frac{d^3 k}{(2\pi)^3} \frac{2 m^f
m^b}{m^\psi}\frac{1}{k^2}\punc{,}
\ee
where $\nu_0$ is the `bare' detuning that appears explicitly in the
action.

We use this expression to write $\nu_0$, which appears within $\Xi^\psi$ in
(\ref{StageinF2}), in terms of $\nu$, and then keep terms only of order $g^2$.
The renormalized expression for $\Delta\Phi$ is then given by
\begin{widetext}
\be
\Delta\Phi = g^2\,\int\frac{d^3 k}{(2\pi)^3} \int\frac{d^3 \ell}{(2\pi)^3}
\left\{\frac{\left[n\sub{F}(\xi^f_{\ell-k})-
n\sub{F}(\xi^\psi_\ell)\right]\left[n\sub{B}(\xi^b_k)-n\sub{B}(\xi^\psi_{\ell} -
\xi^f_{\ell-k})\right]}{\xi^b_k+\xi^f_{\ell-k}-\xi^\psi_{\ell}}
+\; \frac{2 m^f m^b}{m^\psi}\frac{n\sub{F}(\xi^\psi_\ell)}{k^2}\right\}\punc{,}
\ee
where the dispersion relation $\xi^\psi$ now involves the renormalized
(physical) detuning $\nu$. (We have retained the same symbols for the new,
renormalized quantities.)

This expression can be simplified somewhat by performing a further change of
variable, taking $k \rightarrow k + (m^b/m^\psi)\ell$, and also making use of
the result
\be
n\sub{F}(x)n\sub{B}(y-x)+n\sub{F}(y)n\sub{B}(x-y) = -
n\sub{F}(x)n\sub{F}(y)\punc{.}
\ee
We have finally
\be
\Delta \Phi\;= g^2\,\int\frac{d^3 k}{(2\pi)^3} \int\frac{d^3 \ell}{(2\pi)^3}
\frac{1}{\smallfrac{m^\psi}{2m^f m^b}k^2-\nu}
\left[ n\sub{F}(\xi^f)n\sub{B}(\xi^b)-
n\sub{F}(\xi^\psi)n\sub{B}(\xi^b)+n\sub{F}(\xi^f)n\sub{F}(\xi^\psi)-
n\sub{F}(\xi^\psi)\frac{2m^f m^b}{m^\psi}\frac{\nu}{k^2} \right]\punc{,}
\label{FinalDeltaF}
\ee
\end{widetext}
in which the energies $\xi^\psi$, $\xi^f$ and $\xi^b$ should be evaluated at the
following momenta:
\bea
\xi^\psi &\equiv& \xi^\psi(\ell)\\
\xi^f &\equiv& \xi^f(k-\smallfrac{m^f}{m^\psi}\ell)\\
\xi^b &\equiv& \xi^b(k+\smallfrac{m^b}{m^\psi}\ell)\punc{.}
\eea
Note that there is no singularity in the integral over $k$ in
(\ref{FinalDeltaF}), since the numerator also vanishes at the point where
\be
|k|=\sqrt{\nu\frac{2m^f m^b}{m^\psi}}
\ee
(for $\nu > 0$).

The expression for $\Delta\Phi$ must be differentiated with respect to the
chemical potentials to give the correction to the number of each species of
particle. The resulting integral can then be performed numerically.

\begin{figure}
        \resizebox{\columnwidth}{!}{
            \includegraphics{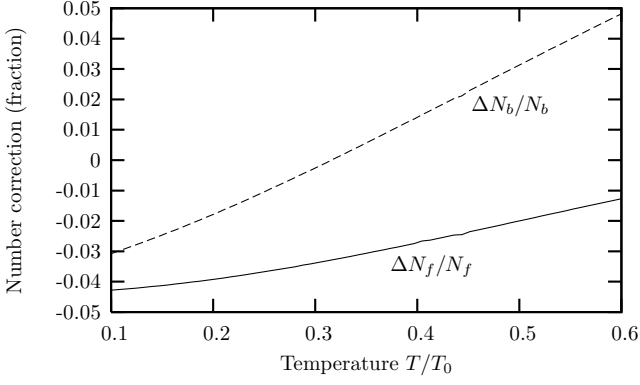}
        }
\caption{\label{Graph1}Two-loop corrections to particle numbers, shown as a
fraction of the total particle numbers to lowest order. The masses of the
particles and their total numbers are the same as in Fig.~\ref{Graph0},
$N_f/N_b = 1.11$. The dimensionless coupling is
$\gamma^2/T_0 = 2.5 \times 10^{-4}$.
At each temperature, the corrections have been evaluated
taking the detuning at its critical value, from Fig.~\ref{Graph0}.}
\end{figure}
The results of such a calculation are shown in Fig.~\ref{Graph1}, where we plot
the corrections to the particle numbers. These have been divided by the total
numbers evaluated using the results of Section
\ref{MFFixedN}. The atomic masses have been taken to be equal, as in
Fig.~\ref{Graph0}, and, at each temperature value, the detuning takes on its
critical value.

We have taken for the dimensionless coupling
$\gamma^2/T_0 = 2.5 \times 10^{-4}$, as in Fig.~\ref{Graph0}, so that this
corresponds to a narrow Feshbach resonance. Since the correction
everywhere is less than $5\%$ of the total numbers, we expect that the
mean-field results provide a good approximation for this case. The
magnitude of the correction scales with $g^2\propto \gamma$, so that the
quantitative predictions become less reliable for a broader resonance.

\section{Low-density approximation}
\label{LowDensityApproximation}

In this section, we develop a low-density approximation involving all orders in
the coupling $g$.

\subsection{Diagrammatic description}

In the mean-field approximation, we have included in the boson self-energy such
diagrams as
\be
\parbox{35mm}{
\begin{picture}(35,20)
\put(0,0){\includegraphics{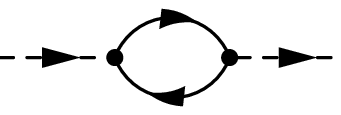}}
\input{fmfPaper.t4}
\end{picture}
}
\ee
whose amplitude is proportional to the density of atoms in the system. (A
fermionic atom must be present in the system initially, in order to couple with
the boson.) Diagrams involving further loops with bosonic or fermionic atoms,
such as
\be
\parbox{35mm}{
\begin{picture}(35,35)
\put(0,0){\includegraphics{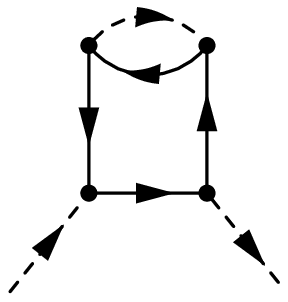}}
\input{fmfPaper.t5}
\end{picture}
}
\ee
are proportional to higher powers of the density.

There is, however, a correction to the molecular propagator of order zero in
density coming from the process shown in (\ref{BubbleDiagram1}), which can take
place in the vacuum. To make a consistent low-density approximation, we must
therefore correct the molecular propagator using a Dyson equation
\be
\label{Dyson}
\parbox{15mm}{
\begin{picture}(15,15)
\put(0,0){\includegraphics{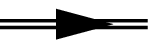}}
\input{fmfPaper.t6}
\end{picture}
}
\quad=\quad
\parbox{15mm}{
\begin{picture}(15,15)
\put(0,0){\includegraphics{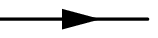}}
\input{fmfPaper.t7}
\end{picture}
}\quad+\quad
\parbox{35mm}{
\begin{picture}(35,20)
\put(0,0){\includegraphics{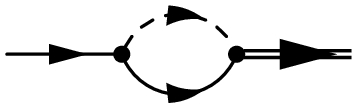}}
\input{fmfPaper.t8}
\end{picture}
}
\ee
where only the term in the bubble diagram (\ref{BubbleDiagram1}) of order zero
in density is to be included.

The boson self-energy diagram
\be
\parbox{35mm}{
\begin{picture}(35,20)
\put(0,0){\includegraphics{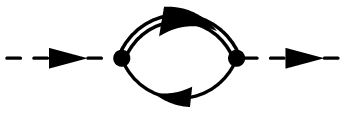}}
\input{fmfPaper.t8}
\end{picture}
}
\ee
should now be used to give a low-density approximation for the phase transition.

\subsection{Calculations}
\label{SectionCalculations}

The Dyson equation (\ref{Dyson}) gives the relation between the reciprocals of
the bare and full Green functions
\be
\Xit^\psi_q = \Xi^\psi_q - \frac{g^2}{\beta} \sum_p G^f_{q-p}G^b_p\punc{;}
\ee
compare (\ref{EffBoseDisp}). [The sign difference results from the fermion loop
in (\ref{FirstGraph}).]

Including only corrections to order zero in density, by dropping the Bose-
Einstein and Fermi-Dirac factors, gives
\be
\Xit^\psi_q = \Xi^\psi_q - g^2 \int \frac{d^3 k}{(2\pi)^3} \left(
\frac{1}{\xi^b_k + \xi^f_{\ell-k}-i\nu_m}-\frac{2m^f m^b}{m^\psi
k^2}\right)\punc{,}
\ee
where $q \equiv (\ell, \nu_m)$. (The second term in the parentheses comes from
renormalizing the detuning $\nu$, as in Section \ref{Renormalize} above.) The
integral can be performed analytically, to give
\be
\Xit^\psi_q = \Xi^\psi_q + 2\gamma\sqrt{\xi^\psi_\ell-\nu-i\nu_m}\punc{,}
\ee
where $\gamma$, defined in (\ref{Definegamma}), has been used.

This function can be continued to one that is analytic everywhere except along
the real axis, by replacing $i\nu_m$ by $z$. In terms of $z$, the full Green
function is
\be
\Gt^\psi_\ell(z) = \frac{-1}{z-\xi^\psi_\ell - 2\gamma \sqrt{\xi^\psi_\ell - \nu
- z}}\punc{.}
\ee
Along the real axis, the square root has a branch cut for $z > \xi^\psi_\ell -
\nu$ which corresponds to the continuum of free-atom excitations. For $\nu < 0$,
$\Gt^\psi$ has a single pole at the real value
\be
z_0 = \xi^\psi_\ell - 2\gamma^2\left(1 - \sqrt{1 -
\smallfrac{\nu}{\gamma^2}}\right)\punc{,}
\ee
corresponding to the renormalized molecule. For $\nu > 0$, there are no poles,
since the molecule has a finite lifetime, decaying into two atoms.

\subsection{Spectral representation}

These analytical properties are best summarized using the spectral
representation for $\Gt^\psi$,
\be
\Gt^\psi_\ell(z) = \int_{-\infty}^{\infty}dx \,\frac{\rho^\psi_\ell(x)}{x -
z}\punc{.}
\ee
The `spectral density' is given by
\begin{widetext}
\be
\rho^\psi_\ell(x) = \Theta(x-\xi^\psi_\ell + \nu) \:
\frac{2\gamma}{\pi}\,\frac{\sqrt{x-\xi^\psi_\ell + \nu}}{{\left(x-
\xi^\psi_\ell\right)}^2 + 4\gamma^2 \left(x-\xi^\psi_\ell +
\nu\right)}
+ \;\Theta(-\nu) \frac{\sqrt{1 - \smallfrac{\nu}{\gamma^2}} - 1}{\sqrt{1 -
\smallfrac{\nu}{\gamma^2}}}\:\delta\!\left(x-\xi^\psi_\ell-
2\gamma^2\left(\sqrt{1 - \smallfrac{\nu}{\gamma^2}} - 1\right)\right)\punc{,}
\label{SpectralDensity}
\ee
\end{widetext}
where $\Theta$ is the unit step function and $\delta$ is the Dirac delta
function.

\subsubsection{Weak-coupling limit}

It should be noted that, in the limit of small coupling, $\gamma\rightarrow 0$,
the first term of (\ref{SpectralDensity}) involves the Lorentzian representation
of the Dirac delta function,
\be
\lim_{\varepsilon\rightarrow 0} \frac{\varepsilon}{\pi(t^2+\varepsilon^2)} =
\delta(t)\punc{.}
\ee
For $\gamma$ small enough, the first term of (\ref{SpectralDensity}) has weight
only near $x = \xi^\psi_\ell$, where $x-\xi^\psi_\ell+\nu$ can be replaced by
$\nu$. The limit of vanishing $\gamma$ is therefore given by
\be
\lim_{\gamma\rightarrow 0} \frac{2\gamma}{\pi}\,\frac{\sqrt{x-\xi^\psi_\ell +
\nu}}{{\left(x-\xi^\psi_\ell\right)}^2 + 4\gamma^2 \left(x-\xi^\psi_\ell +
\nu\right)} = \delta(x - \xi^\psi_\ell)\punc{,}
\ee
so that the spectral density becomes, in this limit,
\bea
\rho^\psi_\ell(x) &\rightarrow& \Theta(x-\xi^\psi_\ell +
\nu)\delta(x - \xi^\psi_\ell) + \Theta(-\nu)\delta(x-\xi^\psi_\ell)\nonumber\\
&=& \delta(x-\xi^\psi_\ell)
\punc{,}
\eea
which is precisely the result for the bare molecule, used in Section
\ref{EffActBoson}.

\subsection{Effective action for bosons}

Using the modified spectral density for the molecule, (\ref{SpectralDensity}),
we can compute the quadratic term in the effective action for the boson field.
Following Section \ref{MF2}, the modified coefficient is
\be
\DeltaLDA = -\mu^b + g^2\int\frac{d^3 \ell}{(2\pi)^3}
\frac{1}{\beta}\sum_{\nu_m} G^f_\ell(i\nu_m)\,\Gt^\psi_\ell(i\nu_m)\punc{.}
\ee
We can now use the spectral representation for both Green functions and then
perform the Matsubara sum. Since the spectral density for the $f$ atom has the
form
\be
\rho^f_\ell(x) = \delta(x - \xi^f_\ell)\punc{,}
\ee
the low-density approximation to the discriminant is given by
\be
\label{LDADelta}
\DeltaLDA = -\mu^b + g^2\int\frac{d^3 \ell}{(2\pi)^3} \,\int_{-
\infty}^{\infty}dx\,\rho^\psi_\ell(x)\frac{n\sub{F}(x)-n\sub{F}(\xi^f_\ell)}{x-
\xi^f_\ell}\punc{,}
\ee
with $\rho^\psi_\ell(x)$ given by (\ref{SpectralDensity}), above. This integral
can be performed numerically.

Since $\DeltaLDA$ provides a correction to $\Delta$, we must find some measure
by which to determine the significance of this correction. A comparison with
$\Delta$ is obviously not possible, since this vanishes everywhere along the
mean-field curve in Fig.~\ref{Graph0}. Instead, we shall find a lowest-order
correction to the critical detuning by evaluating
\be
\label{Definedeltanu}
\delta\nu \equiv \left(\DeltaLDA -
\Delta\right)\Big{/}\parderat{\Delta}{\nu}{\beta,N_f,N_b}\punc{.}
\ee
(Finding the curve $\DeltaLDA = 0$ exactly would require a much larger
computational effort, since $\DeltaLDA$ takes considerably more time to evaluate
than $\Delta$. Instead, $\delta\nu$ provides the first step of a solution of
$\DeltaLDA = 0$ by Newton's iterative method.)

The partial derivative on the right-hand side of (\ref{Definedeltanu}) is given
by
\begin{widetext}
\be
\parderat{\Delta}{\nu}{N_f,N_b} = \quad\parderat{\Delta}{\nu}{\mu^f,\mu^b} +
\parderat{\Delta}{\mu^f}{\mu^b,\nu}\parderat{\mu^f}{\nu}{N_f,N_b}
+ \parderat{\Delta}{\mu^b}{\mu^f,\nu}\parderat{\mu^b}{\nu}{N_f,N_b}\punc{,}
\ee
\end{widetext}
where all the derivatives are to be taken at constant $\beta$. To lowest order
in the coupling, this can be replaced by
\be
\parderat{\Delta}{\nu}{N_f,N_b} = \quad\parderat{\mu^b}{\nu}{N_f,N_b} +
\Order{g^2}\punc{,}
\ee
using (\ref{DiscriminantIntegral}). The quantity on the right-hand side is the
increase in the boson chemical potential needed to compensate an increase in the
detuning, and keep the particle numbers unchanged. This can be evaluated using
the expressions given in Section \ref{MFFixedN}.

\begin{figure}
    \resizebox{\columnwidth}{!}{
        \includegraphics{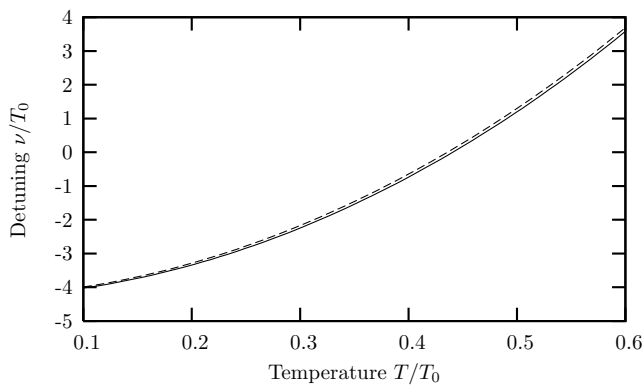}
    }
\caption{\label{Graph2}The phase boundary, without (solid line) and with (dashed
line) the correction to the detuning of Section \ref{LowDensityApproximation}.
The solid line is the phase boundary as in Fig.~\ref{Graph0}, using the same
parameters (and dimensionless coupling $\gamma^2/T_0 = 2.5 \times 10^{-4}$). The
dashed line includes the correction $\delta\nu$, from
(\ref{Definedeltanu}), found using a low-density approximation. For clarity,
this correction has been exaggerated by a factor of $3$. Since the correction is
negligible for these parameters, the mean-field results of Sections
\ref{MFforMu} and \ref{EffActBoson} are sufficient.}
\end{figure}

The results of such a calculation are shown in Fig.~\ref{Graph2}. The solid line
is the phase boundary shown in Fig.~\ref{Graph0}, using the same parameters.
The dashed line is the same curve with the quantity $\delta\nu$ from
(\ref{Definedeltanu}) added. (For clarity, we have multiplied $\delta\nu$ by a
factor of $3$.)
The small magnitude of the correction suggests that the results presented in
Sections \ref{MFforMu} and \ref{EffActBoson} are valid, for the parameters
chosen. In this case, it is therefore unnecessary to use a low-density
approximation; the mean-field result is sufficient.

\section{Conclusions}
\label{sec:conc}

From a theoretical perspective, the main contribution of this
paper is a description of a quantum phase transition with an
intimate connection to the Luttinger theorem. On one side of the
transition (see Fig.~\ref{ZeroTphases}), in the 2 FS, no BEC phase,
there are 2 Fermi surfaces with a separate Luttinger theorem for
each Fermi surface. Remarkably, one Fermi surface is constrained
by the total number of {\em bosons\/} $N_b$, while the other is
controlled by $N_f-N_b$, where $N_f$ is total number of fermions.
These two Luttinger theorems are consequences of two number
operators that commute with the Hamiltonian: the operator
$f^\dagger f - b^\dagger b$ in Eq.~(\ref{const1}), and the
operator $f^\dagger f + \psi^\dagger \psi$ in Eq.~(\ref{const2}).
On the other side of the quantum critical point is the 2 FS + BEC
phase. Here a Bose-Einstein condensate is present, and the
condensate effectively thwarts one of the Luttinger constraints.
The single remaining Luttinger theorem demands only that the total
volume enclosed by both Fermi surfaces is constrained by $N_f$. We
presented a theory for this transition, along with phase diagrams
as a function of system parameters.

It is intriguing to note a connection between the above quantum
phase transition and a seemingly disconnected, recent analysis of
a quantum phase transition in a Kondo lattice model of the heavy
fermion compounds.\cite{ssv} This was a model of electrons
occupying localized $f$ orbitals (the $f_\sigma$ electrons, where
$\sigma$ is a spin index) interacting the itinerant electrons in
the conduction band (the $c_\sigma$ electrons). A boson, $b$, was
introduced to represent the hybridization between the orbitals.
The connection between the Kondo lattice model and the model of
the present paper now becomes clear once we identify the
$c_\sigma$ electrons with the molecular fermionic state $\psi$,
and the $f_\sigma$ electrons with the $f$ fermions. The Kondo
lattice model also has two number constraints analogous to
Eq.~(\ref{const1}) and Eq.~(\ref{const2}), with one crucial
difference: the first constraint is local rather than global, and
applies separately on each lattice site. With this mapping, the
heavy Fermi liquid FL state of Ref.~\onlinecite{ssv} can be
identified with the 2 FS + BEC and the 1 FS + BEC phases. Further,
the FL* phase of Ref.~\onlinecite{ssv} is the analog of the
present 2 FS, no BEC phase. The FL* phase also has two Luttinger
theorems, one fixing the volume of the conduction band Fermi
surface of electronic quasiparticles, and the other the volume of
the `spinon' Fermi surface. The presence of a local rather than a global
constraint implies that there is an additional gauge force that
affects the spinon Fermi surface and the quantum critical
fluctuations of the Kondo lattice. Such gauge forces are absent in
our present considerations of Bose-Fermi mixtures, but, apart from
this absence, there is a remarkable similarity to the FL-FL*
transition in Kondo lattice models.

On the experimental front, an obvious signature of the quantum
phase transition in the Bose-Fermi mixtures is in the evolution of
the Bose-Einstein condensate. It would be interesting to scan the
detuning and look for the disappearance of the condensate fraction
at the lowest temperatures. The corresponding
``superfluid-normal'' transition should also survive at $T>0$,
where its signatures are similar to the $\lambda$ transition in
$^4$He.

A more dramatic, but experimentally less accessible, signature of
the transition lies in the values of the Fermi wavevectors, as
sketched in Fig.~\ref{ZeroTk}. Measuring the Fermi wavevectors
would allow detection of a Fermi surface constrained by the number
of bosons, and its eventual evolution across the transition to a
Fermi surface constrained by the total number of fermions.

Finally, it should be noted that we have not addressed here the
alternative of a paired state of fermions. Since we have dealt
with spin-polarized fermions, $s$-wave pairing between the atoms
is excluded, but $p$-wave pairing remains a possibility.
\cite{gurarie,ho} There is
also the more novel possibility for pairing between the fermionic
atoms $f$ and the molecules $\psi$, which could be favorable when
the two Fermi wavenumbers are approximately equal. This would then
lead to condensation of a composite boson comprised of two fermionic
atoms and one bosonic atom. We intend to investigate this
possibility further in future work.

\acknowledgments We thank R.~Hulet and W.~V.~Liu
for valuable discussions. This research was supported by the
National Science Foundation under grant DMR-0098226, and under
grant DMR-0210790. S.S. was also supported by the John Simon
Guggenheim Memorial Foundation.

\appendix*

\section{Stability against phase separation}
\label{PhaseSeparation}

This section uses the mean-field results of Section \ref{MFforMu}. The
temperature will be taken as zero throughout.

\subsection{The compressibility matrix}

To establish the stability of the system against separation into two coexisting
fluids, we evaluate the compressibility matrix, defined by
\be
\label{Kprimedefined}
K_{\alpha\beta}' = -\frac{\partial^2
\Phi}{\partial\mu^\alpha\partial\mu^\beta}\punc{,}
\ee
for $\alpha,\beta\in\{f,b\}$.

We now define the (canonical) free energy $F(N_f,N_b)$ by a Legendre
transformation,
\be
F(N_f,N_b) = \Phi(\mu^f,\mu^\psi) + \mu^f N_f + \mu^b N_b\punc{,}
\ee
where $N_f$ and $N_b$ are the total number of Fermi and Bose atoms,
respectively. (Note that the full fermion and boson numbers, which are conserved
by the Hamiltonian, are used.) The compressibility matrix $K'$ is then the
inverse of the Hessian of $F$, so that complete stability against phase
separation requires that $K'$ be positive semidefinite.

It is in fact easier to work with the matrix $K_{\alpha\beta}$, given by the
same expression (\ref{Kprimedefined}), but with $\alpha,\beta\in\{f,\psi\}$.
This amounts to a simple (but not orthogonal) change of basis; it is sufficient
(and necessary) for $K'$ to be positive semidefinite that $K$ be the same.

We begin with (\ref{FreeEnergy}) and (\ref{DefineR}) and use (\ref{EqCondensate})
to determine the implicit dependence of $\varphi$ on the chemical potentials.
We must then take second derivatives with respect to the two chemical potentials
to find the compressibility matrix. In the presence of a condensate, this leads
to an expression
\be
K_{\alpha\beta} = K_{\alpha\beta}^{(0)} + \frac{r_\alpha r_\beta}{\lamt}\punc{,}
\ee
where $K_{\alpha\beta}^{(0)}$ is the matrix of second derivatives, evaluated at
fixed $\varphi$ and $r_\alpha$ is a function whose form will not concern us here.

The denominator of the second term is
\be
\label{resultantinteraction}
\lamt = \lambda + g^4\int_{k_0^F}^{k_0^\Psi}\frac{dn(k)}{W_k^3}
- \half g^4 \left(z^F + z^\Psi\right)\punc{,}
\ee
where
\be
\label{Definez}
z^x = {\left.\frac{m^\psi m^f dn/dk}{k\left(\xi^f_k +
\xi^\psi_k\right)\left(m^f\xi^f_k +
m^\psi\xi^\psi_k\right)}\right|}_{k=k_0^x}
\ee
and
\be
W_k = \sqrt{{\left(\xi^f_k-\xi^\psi_k\right)}^2 + 4g^2\varphi^2}\punc{.}
\ee
When $\lamt$ goes through zero, the determinant of $K$ diverges, so that the
Hessian of $F$ becomes singular, signifying that one of its eigenvalues vanishes.
This marks the onset of instability; we conclude that stability requires that
$\lamt > 0$.

When there is no condensate, \ie\ in the phase labeled `2 FS, no BEC' in
Fig.~\ref{ZeroTphases}, it is found that the system is always stable.

\subsection{Physical interpretation}

The obvious physical interpretation of $\lamt$ is that it represents the resultant
interaction between the bosons, coming partly from the explicit term $\lambda$ in
the Hamiltonian (\ref{Hamiltonian}) and partly from the interaction induced by
coupling to the fermions. This induced interaction can alternatively be found
directly by continuing the expansion (\ref{QuadAction}) to fourth order in $b$
and $\bar{b}$.

A resultant interaction of the form (\ref{resultantinteraction}) is familiar
from the case where the molecular degrees of freedom are not included explicitly
in the Hamiltonian.\cite{Buechler,Viverit} This corresponds to our model for
$\nu\gg 0$, when only virtual molecules are formed and the coupling
term $\psi^\dagger f b$ in the Hamiltonian (\ref{Hamiltonian}) can be replaced by
a boson-fermion scattering of the form $f^\dagger b^\dagger b f$. The induced
interaction then comes from the diagram
\be
\parbox{35mm}{
\begin{picture}(35,20)
\put(0,0){\includegraphics{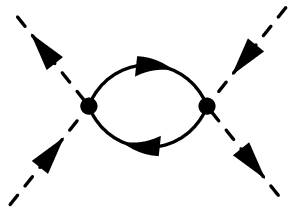}}
\input{appendix.t1}
\end{picture}
}
\punc{,}
\label{Fish}
\ee
which gives a term proportional to the density of states at the Fermi surface (at
$T = 0$).

In this case, the induced interaction is always attractive, as can be shown by a
simple physical argument. For experimentally accessible parameters, however, it
is not strong enough to overcome the intrinsic repulsion between the bosons, so
that the phase is stable.\cite{Viverit} In our notation, the boson-fermion
scattering is suppressed by a factor of $1/\nu$, so that the induced interaction
falls off as $1/\nu^2$. For $\nu\ll 0$, a similar picture is obtained, with the
atomic and molecular fermions exchanging r\^oles.

In the case of intermediate $\nu$, the induced interaction is no longer so heavily
suppressed, but it is also no longer the case that it is always attractive. The
physical picture is clarified in this case by rewriting the action in
(\ref{eqAction}) in terms of the fermions $F$ and $\Psi$ introduced in Section
\ref{MFHamiltonian}. These fermions are defined so that there is no coupling term
in the action linear in $\varphi = \Mean{b}$; instead, the lowest order interactions
have the form $\bar{F}\,\varphi^2\,F$ and $\bar{F}\,\varphi^4\,F$,
and the same for $\Psi$. The former reproduces exactly the diagram (\ref{Fish})
above, with $f$ replaced by $F$ and $\Psi$: physically this is a boson-fermion
scattering inducing an attractive interaction between the bosons, as described
above. This accounts for the final term in (\ref{resultantinteraction}). Note that
the exclusion principle requires the momenta of the two fermion lines to be exactly
at the Fermi surface, leading to $z^x$ being evaluated at $k_0^x$.

The term $\bar{F}\,\varphi^4\,F$ produces the diagram
\be
\parbox{25mm}{
\begin{picture}(25,25)
\put(0,0){\includegraphics{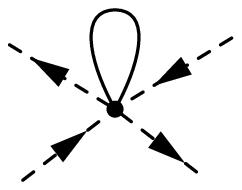}}
\input{appendix.t2}
\end{picture}
}
\label{Spider}
\ee
which also represents an induced boson-boson interaction and accounts for the
integral in (\ref{resultantinteraction}). Since $W_k \ge 0$, it is always repulsive
and represents the fact that the fermion energy is lowered by a uniform
distribution of bosons.

\subsection{Results}

\begin{figure}
    \resizebox{\columnwidth}{!}{
        \includegraphics{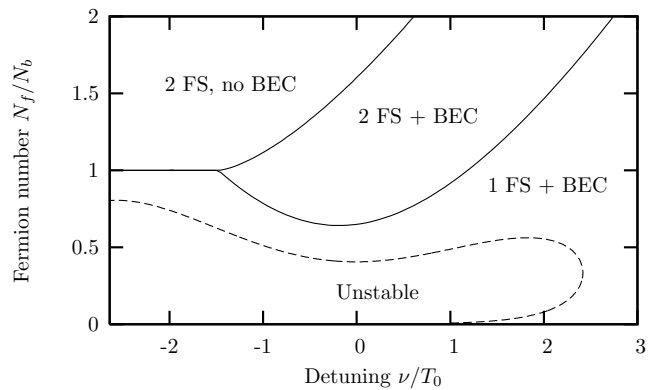}
    }
\caption{\label{ZeroTstability}The phase diagram at $T = 0$, as in
Fig.~\ref{ZeroTphases}, with couplings $\gamma^2/T_0 = 2.0\times 10^{-2}$ and
$\lambda^2 (m^b)^3 T_0 = 2\times 10^{-3}$.
The other parameters, and the labels for the three phases, are the same as
in Fig.~\ref{ZeroTphases}. The region where the phase is unstable, as determined
in the Appendix, is indicated.}
\end{figure}

The sign of the resultant interaction $\lamt$ must be calculated numerically to
determine whether the system is indeed stable. Using the parameters from
Fig.~\ref{ZeroTphases}, stability is found everywhere within the plot for cases
(a) and (b). In case (c), where the coupling $g$ is larger relative to $\lambda$,
there is a region of the diagram where the phase is not stable; this is shown
in Fig.~\ref{ZeroTstability}.

For large $|\nu|$ the attractive coupling from the diagram (\ref{Fish}) is
suppressed by a factor $1/\nu^2$ as described above, so that the system becomes
stable. (The region for large negative $\nu$ is not visible on this plot.) For
intermediate values of $|\nu|$, the induced coupling becomes larger than the
intrinsic coupling, $\lambda$, and it is the competition between the two diagrams
(\ref{Fish}) and (\ref{Spider}) that determines the stability.

Stability is therefore favored by a higher $N_f/N_b$, since this increases
$k_0^\Psi$ and hence the phase space for the diagram (\ref{Spider}). The other
diagram, (\ref{Fish}), increases more slowly with $k_0^\Psi$ since the internal
fermion lines are restricted to be at the Fermi surface. For intermediate $|\nu|$
and very small $N_f/N_b$, on the order of $10^{-3}$, the instrinsic interaction once
more dominates the induced and the system is stable. This region is too small to
be seen in Fig.~\ref{ZeroTstability}.

An analysis similar to that carried out in Ref.~\onlinecite{Viverit} could be
performed to determine the stabilities of the alternative, mixed phases. It should
be noted, however, that, as can be seen in Fig.~\ref{ZeroTstability}, the
boundaries between the three phases are not disturbed
at the parameters we have considered.

Furthermore, the analysis above shows that increasing the coupling
$g$ (or equivalently $\gamma^2/T_0$) beyond the value used in
Fig.~\ref{ZeroTstability} would increase the value of $|\nu|$ required
for stability at small $N_f/N_b$ (\ie\ extend the unstable region
to larger $|\nu|$), but would not decrease the stability at
intermediate $|\nu|$. This follows from the fact that the latter
is determined by the competition between the two diagrams
(\ref{Fish}) and (\ref{Spider}), whose relative magnitude does not
depend on $g$. We therefore expect that, for a broad Feshbach resonance,
there remains a large region of stability for intermediate values of
$|\nu|$, similar to that in Fig.~\ref{ZeroTstability}.


\begin{thebibliography}{99}

\bibitem{stwalley1976} W.\ C.\ Stwalley, Phys.\ Rev.\ Lett.\ {\bf 37}, 1628 (1976).

\bibitem{tiesinga1993} E.\ Tiesinga, B.\ J.\ Verhaar, and H.\ T.\ C.\ Stoof,
                       Phys.\ Rev.\ A {\bf 47}, 4114 (1993).

\bibitem{jochim2003} S.\ Jochim, M.\ Bartenstein, A.\ Altmeyer, G.\ Hendl, S.\ Riedl,
                     C.\ Chin, J.\ Hecker Denschlag, and R.\ Grimm,
                     Science {\bf 302}, 2101 (2003).

\bibitem{greiner2003} M.\ Greiner, C.\ A.\ Regal, and D.\ S.\ Jin,  Nature (London) {\bf 426}, 537
(2003).

\bibitem{zwierlein2003} M.\ W.\ Zwierlein, C.\ A.\ Stan, C.\ H.\ Schunck,
                        S.\ M.\ F.\ Raupach, S.\ Gupta, Z.\ Hadzibabic, and W.\ Ketterle,
                         Phys.\ Rev.\ Lett.\ {\bf 91}, 250401 (2003).

\bibitem{leo}  L.\ Radzihovsky, J.\ Park, and P.\ B.\ Weichman, Phys.\ Rev.\ Lett.\ {\bf 92}, 160402 (2004).

\bibitem{ss} M.\ W.\ J.\ Romans, R.\ A.\ Duine, S.\ Sachdev, and H.\ T.\ C.\ Stoof,
Phys.\ Rev.\ Lett.\ {\bf 93}, 020405 (2004).

\bibitem{truscott} A.\ G.\ Truscott, K.\ E.\ Strecker, W.\ I.\ McAlexander,
G.\ B.\ Partridge, and R.\ G.\ Hulet, Science {\bf 291}, 2570 (2001).

\bibitem{schreck} F.\ Schreck, L.\ Khaykovich, K.\ L.\ Corwin, G.\ Ferrari, T.\ Bourdel,
J.\ Cubizolles, and C.\ Salomon, Phys.\ Rev.\ Lett.\ {\bf 87}, 080403
(2001).

\bibitem{stan} C.\ A.\ Stan, M.\ W.\ Zwierlein, C.\ H.\ Schunck, S.\ M.\ F.\ Raupach, and
W.\ Ketterle, Phys.\ Rev.\ Lett.\ {\bf 93}, 143001 (2004).

\bibitem{inouye} S.\ Inouye, J.\ Goldwin, M.\ L.\ Olsen, C.\ Ticknor, J.\ L.\ Bohn, and
D.\ S.\ Jin, Phys.\ Rev.\ Lett.\ {\bf 93}, 183201 (2004).

\bibitem{fwgf} M.\ P.\ A.\ Fisher, P.\ B.\ Weichman, G.\ Grinstein, and
D.\ S.\ Fisher, Phys.\ Rev.\ B {\bf 40}, 546 (1989).

\bibitem{book} S.\ Sachdev, {\em Quantum Phase Transitions},
Cambridge University Press, Cambridge (1999).

\bibitem{Yabu}
H.\ Yabu, Y.\ Takayama, and T.\ Suzuki, Physica B {\bf 329-333}, 25
(2003).

\bibitem{Duine}
R.\ A.\ Duine and H.\ T.\ C.\ Stoof, Phys.\ Rep.\ {\bf 396}, 115 (2004).

\bibitem{ohashi:130402}
Y.\ Ohashi and A.\ Griffin, Phys.\ Rev.\ Lett.\ {\bf 89}, 130402
(2002).

\bibitem{luttingerward} J.\ M.\ Luttinger and J.\ C.\ Ward,
Physical Review {\bf 118}, 1417 (1960).

\bibitem{potthoff} M.\ Potthoff, cond-mat/0406671 (2004).

\bibitem{agd} A.\ A.\ Abrikosov, L.\ P.\ Gorkov, and I.\ E.\ Dzyaloshinski,
{\em Methods of Quantum Field Theory in Statistical Physics}, Dover Publications Inc.,
New York (1975).

\bibitem{hertz} J.\ A.\ Hertz, Phys.\ Rev.\ B {\bf 14}, 1165 (1976).

\bibitem{millis} A.\ J.\ Millis, Phys.\ Rev.\ B {\bf 48}, 7183 (1993).

\bibitem{ssv} T.\ Senthil, S.\ Sachdev, and M.\ Vojta, Phys.\ Rev.\ Lett.\ {\bf 90},
216403 (2003); T.\ Senthil, M.\ Vojta, and
S.\ Sachdev, Phys.\ Rev.\ B {\bf 69}, 035111 (2004).

\bibitem{gurarie} V.\ Gurarie, L.\ Radzihovksy, and A.\ V.\ Andreev, cond-mat/0410620 (2005).

\bibitem{ho} T.-L.\ Ho and R.\ B.\ Diener, cond-mat/0408468 (2004).

\bibitem{Viverit} L.\ Viverit, C.\ J.\ Pethick, and H.\ Smith, Phys.\ Rev.\ A {\bf 61},
053605 (2000).

\bibitem{Buechler} H.\ P.\ B\"uchler and G.\ Blatter, Phys.\ Rev.\ A {\bf 69},
063603 (2004).

\end{thebibliography}
\end{document}